\documentclass[manuscript,nonacm]{acmart}
\AtBeginDocument{%
  }

\setcopyright{acmlicensed}
\copyrightyear{2026}
\acmYear{2026}
\acmDOI{XXXXXXX.XXXXXXX}
\acmConference[Conference acronym 'XX]{Make sure to enter the correct
  conference title from your rights confirmation email}{June 03--05,
  2026}{Woodstock, NY}
\acmISBN{978-1-4503-XXXX-X/2018/06}

\usepackage{booktabs, tabularx, ragged2e}
\usepackage{xcolor}
\usepackage{tikz}
\usepackage{tabularx}
\usepackage{colortbl}
\usepackage{multirow}

\definecolor{LightGray}{rgb}{0.929,0.929,0.929}
\definecolor{HumanBG}{HTML}{E9F2F5}
\definecolor{AIBG}{HTML}{FEEEE8}
\definecolor{TableHeader}{RGB}{225,230,235}
\definecolor{TableShade}{RGB}{246,247,248}

\newcommand{\conditionlabel}[2]{%
  \tikz[baseline=(label.base)]{
    \node[
      fill=#1,
      draw=none,
      rounded corners=1.5pt,
      inner xsep=0.25em,
      inner ysep=0.10ex,
      outer sep=0pt,
      text=black,
      font=\bfseries
    ] (label) {#2};
  }%
}

\newcommand{\humanlabel}[1]{%
  \conditionlabel{HumanBG}{#1}%
}

\newcommand{\ailabel}[1]{%
  \conditionlabel{AIBG}{#1}%
}

\begin{document}

\title[How Knowledge Workers Share Their Expertise with AI Systems]{When My Skill Becomes Agent Skill: How Knowledge Workers Share Their Expertise with AI Systems}

\author{Tianqi Song}
\affiliation{%
  \institution{National University of Singapore}
  \country{Singapore}
  }
\email{tianqi\_song@u.nus.edu}

\author{Zicheng Zhu}
\affiliation{%
  \institution{National University of Singapore}
  \country{Singapore}
}
\email{zicheng@u.nus.edu}

\author{Hancheng Cao}
\affiliation{%
  \department{Goizueta Business School}
  \institution{Emory University}
  \city{Atlanta}
  \country{USA}
}
\email{hancheng.cao@emory.edu}

\author{Yi-Chieh Lee}
\affiliation{%
  \institution{National University of Singapore}
  \country{Singapore}
}
\email{yclee@nus.edu.sg}

\renewcommand{\shortauthors}{Tianqi Song, Zicheng Zhu, Hancheng Cao, Yi-Chieh Lee}

\begin{abstract}

Organizations have long sought to make workers’ expertise reusable by others. 
Agentic AI changes the nature of such reuse by enabling AI systems to act on workers’ knowledge with limited human involvement.
This raises the question of how this shift shapes workers' willingness to share their expertise. 
We conducted an experiment with knowledge workers who created materials incorporating domain knowledge and decided whether to authorize human or AI reuse. 
We find that AI and human reuse differ primarily in \textit{whether} workers choose to share their knowledge, rather than in \textit{what} they choose to share.
Participants' reasoning shifts from prosocial considerations when sharing with humans toward concerns about loss of control, replacement, and downstream governance when sharing with AI.
These findings suggest that AI reuse may intensify tensions between organizational knowledge reuse and contributors’ interests.
We discuss implications for workplace AI and knowledge management policies that preserve workers’ rights and agency.



\end{abstract}

\ccsdesc[500]{Do Not Use This Code~Generate the Correct Terms for Your Paper}
\ccsdesc[300]{Do Not Use This Code~Generate the Correct Terms for Your Paper}
\ccsdesc{Do Not Use This Code~Generate the Correct Terms for Your Paper}
\ccsdesc[100]{Do Not Use This Code~Generate the Correct Terms for Your Paper}

\keywords{Knowledge Sharing, Knowledge contribution,  Agentic AI, Future of Work}




\begin{teaserfigure}
\centering
  \includegraphics[width=\textwidth]{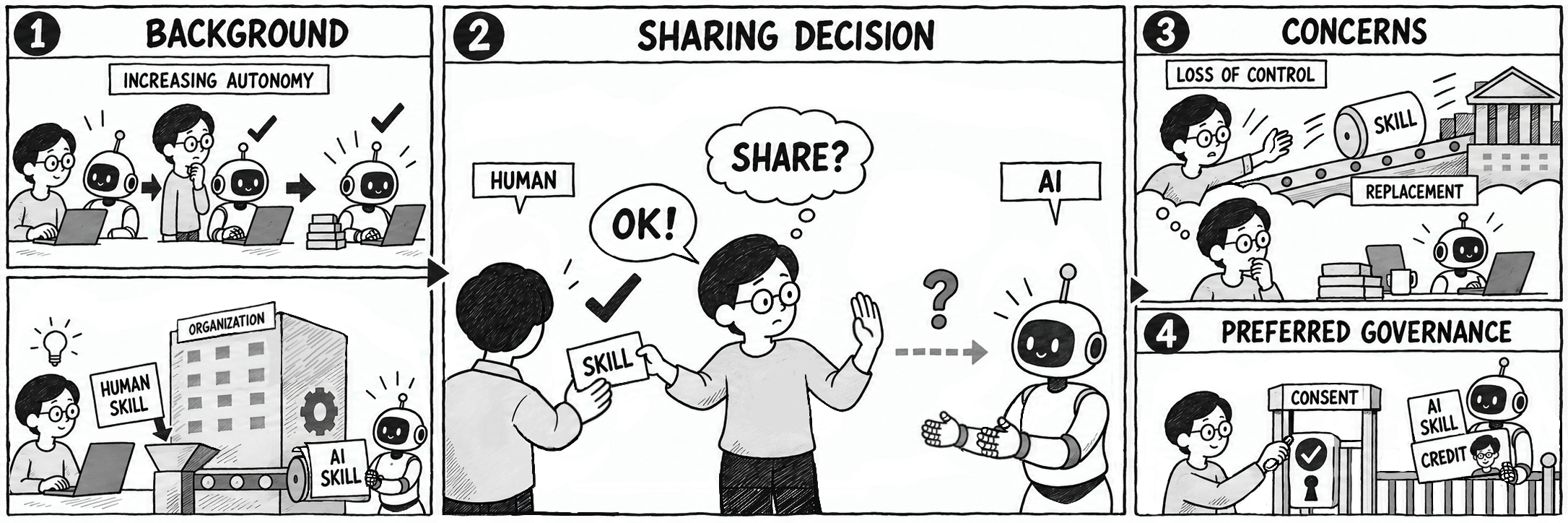}
  \caption{Visual overview of the study. (1) As AI agents become increasingly autonomous, organizations seek to transform workers’ expertise into reusable AI agent skills. (2) When faced with an actual sharing decision, workers were less willing to share their skills with AI than with humans. (3) Their responses highlighted concerns about loss of control and replacement, alongside preferences for governance centered on consent and credit.}
  \label{fig:teaser}
\end{teaserfigure}

\maketitle


\section{Introduction}

Organizational knowledge has long been recognized as a critical source of productivity, innovation, and competitive advantage~\cite{asrar2016systematic, nonaka2007knowledge, bock2005behavioral}.
Much of this knowledge resides in workers' experience and professional judgment, making it difficult for organizations to retain when employees leave or change roles~\cite{kogut1992knowledge, grant1996toward}.
Organizations have therefore relied on documentation, training, mentoring, and knowledge-management systems to preserve workers' expertise and enable its reuse~\cite{argote2003managing, cristache2025influence}.
These practices help make individual expertise available to colleagues and future employees, supporting the continuity and development of organizational knowledge.


Generative and agentic AI introduce new possibilities and challenges for how organizations capture and reuse this expertise~\cite{benderoth2025socially, yuan2026ai}.
Workers' task outputs, reasoning, and workplace interactions can serve as resources for AI systems to perform subsequent tasks.
For example, emerging systems package workers' practices and decision heuristics into reusable \textit{``agent skills''} that AI systems can draw on when performing similar work~\cite{ling2026agent, zhou2026colleague}.
Consider a worker whose explanation of a difficult workplace decision is incorporated into such a skill and applied without their continued involvement.
The worker might welcome its potential to help others handle similar situations more efficiently, yet hesitate to include sensitive conversations or allow the resulting skill to be applied in unfamiliar contexts.
This tension raises a question about \textit{how knowledge workers perceive and approach sharing their expertise for AI reuse.}

Understanding such sharing requires attention to both the contexts in which knowledge is reused and the social relationships through which it is contributed.
Knowledge-reuse research shows that knowledge can be difficult to interpret or apply when producers and reusers do not share the same context~\cite{markus2001toward}.
For contributors, this raises questions about whether others will use their knowledge appropriately and what they are
willing to make available for reuse.
Social-exchange perspectives further explain how contributions are shaped by anticipated reciprocity, recognition, and personal costs~\cite{kankanhalli2005contributing, bock2005behavioral}.
Together, these perspectives suggest that willingness to share depends on both the anticipated uses of knowledge and the balance between collective benefits and individual costs~\cite{cabrera2002knowledge}.

AI reuse may change how workers weigh these considerations when deciding whether to share their knowledge. 
Even workers who value AI and wish to improve its performance~\cite{dell2023navigating} may question whether it can apply their knowledge with appropriate sensitivity to context
~\cite{castelo2019task,longoni2019resistance}.
Contributing expertise may also create tension between improving workplace efficiency~\cite{dell2023navigating} and protecting one's professional role~\cite{chhibber2025will}.
These questions are becoming increasingly consequential as organizations use workers' activities and expertise to develop AI capabilities.
For example, employee backlash against Meta's initiative to collect work activity for AI training illustrates how such efforts can become contested.\footnote{``Meta Is Tracking Employee Keystrokes, Clicks---Causing Backlash.''
\url{https://www.entrepreneur.com/business-news/meta-will-track-every-click-and-keystroke-employees-make-backlash}}
Understanding workers' perspectives is therefore relevant both to supporting knowledge contribution and to establishing boundaries that workers consider acceptable.

Research on AI in organizational knowledge management has largely focused on how AI can support the capture and reuse of workers' expertise~\cite{benderoth2025socially, yuan2026ai, zhou2026colleague}.
A smaller body of emerging work has begun to examine workers' perspectives on contributing knowledge to AI, identifying motivations such as learning opportunities~\cite{renggli2024or} 
and modeling how giving workers collective ownership of their data can encourage knowledge sharing~\cite{cullen2026labor}.
However, less is understood about how workers translate these considerations into explicit authorization decisions over concrete materials derived from their own work: whether they permit reuse, which materials they release, and what conditions they expect to govern subsequent use.
This leads to our central question: \textit{How do knowledge workers decide whether and what to share with AI systems, and how do they understand the consequences and conditions of that sharing?}

To investigate the question, we conducted a mixed-methods study in which 61 full-time professionals completed simulated workplace tasks.
Participants were randomly assigned to share their knowledge with either AI systems or human professionals as the intended recipient, while the task structure and knowledge-extraction process remained consistent across conditions.
Participants first completed an occupation-relevant task designed to elicit their domain knowledge through task outputs, written reasoning, and conversations with simulated coworkers. They then reviewed an LLM-generated skill preview containing knowledge extracted from these materials.
Next, participants were presented with a scenario in which their materials and extracted knowledge could be contributed to an organizational repository for reuse by their assigned recipient. 
They decided whether to authorize reuse and, if so, which categories of materials to release.
Finally, survey ratings and open-ended responses captured participants' anticipated consequences of sharing and their preferred conditions for subsequent use.


Our findings show that participants were less likely to authorize organizational knowledge reuse by AI systems than by human professionals. 
At the category level, differences in release rates between conditions were not statistically significant.
Among participants who authorized AI reuse, conversations were less often shared than task outputs or written reasoning.
Qualitative responses linked these sharing boundaries to concerns about sensitive information and contextual interpretation.
Post-study survey further revealed that participants in in the AI condition anticipated greater loss of expert influence and control than those in the human condition.
Continued control and credit were also expected to increase willingness to share more with AI than with humans, while open-ended responses revealed varied preferences for acknowledgment and anonymity.

This work aims to make three contributions to HCI:
\begin{itemize}
    \item We show that knowledge workers are less likely to authorize reuse of their expertise by AI systems than by human professionals.
    This extends knowledge-sharing research to AI reuse by providing empirical evidence of how workers authorize the use of materials derived from their own work.
    
    \item We provide suggestions for organizations seeking to encourage knowledge sharing with AI. By identifying workers' expectations for recognition and continued control, we highlight ways to support voluntary participation while accommodating different preferences for credit and anonymity.
    
    \item We identify implications for system design and organizational governance that protect workers' interests as their expertise becomes a resource for AI capabilities. Drawing on concerns about expert influence, interpretive control, and professional replacement, we highlight the need for ongoing contributor agency and safeguards against adverse downstream uses.
\end{itemize}



\section{Related Work}

\subsection{Organizational Knowledge Management}
Organizations have long sought to preserve and transfer workers’ knowledge across individuals, teams, and time, as organizational knowledge is a critical driver of performance and competitive advantage in the knowledge economy~\cite{wang2012knowledge}.
When experienced employees change roles or leave, organizations risk losing not only documents and formal procedures, but also the judgment required to recognize exceptions, navigate relationships, and adapt routines to particular situations. 
Knowledge management addresses this problem by examining how organizational knowledge is created, retained, transferred, and reused~\cite{davenport1998working, argote2003managing}. 
A central strategy is knowledge codification: representing workers' experience and expertise in forms that can be stored and made available beyond the situations in which they were originally developed~\cite{schulz2001codification}.

Yet describing work is not the same as knowing how to perform it~\cite{polanyi2009tacit, nonaka2000seci}. 
Ryle's distinction between \textit{``knowing-that''} and \textit{``knowing-how''} suggests that workers may know how to perform a task without being able to state everything their performance involves~\cite{ryle1945knowing}. 
Research on situated learning further shows that such knowledge is embedded in practices, tools, relationships, and organizational settings~\cite{lave1991situated, brown1991organizational}. 
HCI and CSCW research further brings a human-centered perspective to knowledge-sharing systems, by examining how people work, communicate, and develop shared understanding in practice, and how repositories and communication tools support these
activities~\cite{schmidt2012trouble,ackerman2013sharing}.
Taken together, codification can make knowledge more portable, but systems must also support access to the context and expertise needed to use it.

Moreover, whether codified knowledge can be successfully reused also depends on who produced it, who draws on it, and the relationship between their work contexts.
Markus's knowledge-reuse framework distinguishes situations in which people reuse knowledge from their own shared work, draw on others' similar experiences, seek expertise they lack, or repurpose knowledge for new uses~\cite{markus2001toward, markus2002design}.
These situations create different needs for explanation and contextual support, suggesting that making knowledge available does not by itself ensure that others can apply it appropriately.

\subsection{Knowledge Contribution to AI in Organizations}

Generative AI is increasingly used to support workplace tasks and improve productivity~\cite{noy2023experimental, brynjolfsson2025generative}, changing how knowledge is shared and used at work.
Researchers have examined how its presence shapes knowledge sharing and hiding among employees~\cite{li2025synergy, liu2025robot, xu2026linking}, as well as how workers learn from AI systems and apply what they learn~\cite{schut2025bridging, shi2026models}.

Knowledge also flows in the other direction: workers' own expertise can become a resource for AI.
Socially interactive agents and AI expert twins, for example, elicit and preserve knowledge grounded in professional experience~\cite{benderoth2025socially, yuan2026ai}.
Once captured, this knowledge can guide AI systems in carrying out subsequent tasks.
Agent skills provide one way to put this knowledge into practice by bringing together instructions, examples, scripts, and resources for recurring workflows~\cite{ling2026agent, li2026skillsbench}.
AI companies such as OpenAI~\cite{openai2026skills}, Anthropic~\cite{anthropicSkills}, and Google~\cite{googleGeminiSkills} offer tools and guidance for creating such skills, helping users turn established ways of working into reusable instructions for AI.
These instructions can describe general procedures, but they can also reflect how a particular worker approaches a task.
COLLEAGUE.SKILL, for example, draws on an individual's workplace materials to distill their practices and decision heuristics into an executable skill~\cite{zhou2026colleague}.
A worker's expertise may therefore continue to guide tasks without their continued involvement, raising the question of whether they are willing to share what they know for this kind of reuse.

While much of the work on knowledge contribution to AI focuses on building systems that capture and apply expertise, less attention has been paid to how workers decide whether to make their knowledge available for such use. A few studies examine this process from workers' perspectives. For example, Renggli and W\"afler identify motivations and barriers to providing knowledge to AI through exploratory interviews~\cite{renggli2024or}. Cullen et al. combine worker surveys with a field experiment to document gaps between workers' expertise and what organizations have captured, and examine how career incentives affect knowledge provision for AI training~\cite{cullen2026labor}. 
While these studies show how workers' interests and circumstances shape their contributions, they leave open how workers translate their willingness to contribute into authorization decisions over specific materials from their own work. 
A worker may be willing to share a task output while withholding the reasoning or conversations behind it. Thus, the question is both whether workers permit reuse and what they agree to release, as well as how these decisions differ when the intended reuser is an AI system rather than another professional.

\subsection{Knowledge Sharing and Worker Agency}
\subsubsection{Knowledge Contribution as a Social Dilemma}


Although organizations benefit when workers codify and share what they know, doing so requires individual time and effort, while the resulting benefits may accrue mainly to colleagues or the organization. 
Knowledge sharing has therefore been described as a \textit{social dilemma}: collective performance depends on individual contributions that may not serve each worker's immediate interests~\cite{cabrera2002knowledge}.

Prior research explains this decision through both social and individual considerations~\cite{kankanhalli2005contributing, wasko2005should}.
For example, social-exchange theory suggests that people contribute when they expect benefits such as reciprocity, reputation, stronger relationships, or the satisfaction of helping others~\cite{kankanhalli2005contributing}. 
Studies of electronic knowledge repositories similarly identify enjoyment in helping, knowledge self-efficacy, organizational support, and perceived personal costs as important predictors of contribution~\cite{kankanhalli2005contributing, bock2005behavioral, hung2011influence}. 
From this perspective, knowledge sharing is not simply a matter of possessing useful information, 
but also on how workers weigh their contributions against benefits to others and expected returns. 
AI reuse brings this social dilemma into focus by potentially changing who benefits from workers’ knowledge and what contributors receive in return.

\subsubsection{Worker Agency in Knowledge Contribution to AI}
HCI research has increasingly examined how workplace AI reshapes workers’ participation in and relationship to work, including workers’ sense of inclusion, control, and ownership in AI-mediated communication; meaningfulness and accountability in human–AI collaboration; and the interpersonal and organizational implications of workplace AI agents~\cite{kadoma2024role, sadeghian2024soul, hu2026your}. Much of this work takes AI capabilities as an existing feature of the workplace. A related but less examined question concerns how those capabilities are created, particularly when they are built from workers’ knowledge. This shifts attention to workers’ perceived rights and control over whether and how their expertise is used.


\subsection{Research Questions}
This study seeks to understand how knowledge workers perceive and approach sharing their expertise for AI reuse.
Prior research shows how AI systems can capture workers' expertise and turn it into reusable capabilities~\cite{benderoth2025socially, yuan2026ai, zhou2026colleague}.
Yet making expertise reusable also requires understanding workers' willingness to contribute, including what they expect to gain or lose through sharing~\cite{kankanhalli2005contributing, renggli2024or, cullen2026labor}.
When their knowledge can be used beyond their direct involvement, questions of control and ownership discussed in workplace AI research also become relevant~\cite{kadoma2024role}.
Together, these perspectives guide our examination of what workers agree to share, how they understand the potential consequences, and what conditions would make them more willing to contribute.
We address these questions through three research questions:

\begin{itemize}
    \item[\textbf{RQ1}] Do knowledge workers authorize AI systems to reuse their professional knowledge, and what do they agree to share?
\end{itemize}

This question examines both whether workers permit reuse and the boundaries they place on their contributions.
By asking workers to review materials derived from their own work, we examine their choices about what to release rather than willingness to contribute in general.

\begin{itemize}
    \item[\textbf{RQ2}] What benefits and risks do knowledge workers anticipate when sharing their professional knowledge with AI systems?
\end{itemize}

This question examines how workers understand what may happen once their knowledge is shared.
Building on research on knowledge contribution and reuse~\cite{kankanhalli2005contributing, markus2001toward}, we consider both the benefits of making expertise available to others and concerns about its subsequent use and consequences for the contributor.

\begin{itemize}
    \item[\textbf{RQ3}] What governance conditions do knowledge workers expect to change their willingness to share professional knowledge with AI systems?
\end{itemize}

This question turns from the anticipated consequences of sharing to the conditions under which workers would be more willing to contribute.
We examine their preferences for recognition, control over knowledge use, and safeguards concerning downstream consequences.

\section{Methods}


We conducted a mixed-methods study combining a between-subjects experiment, quantitative survey measures, and qualitative open-ended questions. The study received approval from our university's ethics review committee before data collection began.

\subsection{Task Setup}

\begin{table*}[t]
\centering
\caption{Overview of the four occupational simulations used in the study.}
\label{tab:task-design}

\small
\setlength{\tabcolsep}{7pt}
\renewcommand{\arraystretch}{1.22}

\begingroup
\arrayrulecolor{black!55}

\begin{tabularx}{\textwidth}{
    >{\raggedright\arraybackslash}p{0.12\textwidth}
    >{\raggedright\arraybackslash}p{0.18\textwidth}
    >{\raggedright\arraybackslash}X
}
\toprule

\rowcolor{TableHeader}
\textbf{Occupation}
& \textbf{Participant task}
& \textbf{Scenario description} \\

\midrule

Human resources
& Employee Concern Intake
& 
Participants handled an employee's private complaint about their manager, which the employee asked not be filed or disclosed. They decided how to proceed without ruling on the complaint.
\\

\midrule
Marketing
& Campaign Performance Assessment
& Participants evaluated an ongoing campaign using incomplete and partly conflicting evidence, then recommended whether to stop, continue, or modify the campaign. \\

\midrule

Sales
& Renewal Discount Negotiation
& Participants negotiated a customer renewal in which the requested price reduction exceeded their approval limit and could set a precedent for
  other accounts. \\

\midrule
Management
& Performance Review Rating
& Participants assigned an annual performance rating while considering
  evidence that part of the employee's shortfall resulted from their own
  staffing and scheduling decisions. \\

\bottomrule
\end{tabularx}

\endgroup
\end{table*}

Work simulations have been used in organizational behavior research~\cite{allen2026well,trang2021examining,whetzel2014basket} and in assessment contexts, such as e-tray exercises~\footnote{E-tray exercises are online job simulations in which participants handle, prioritize, and respond to workplace emails and documents within a set time.}, to elicit job-relevant behavior and make workers’ application of domain knowledge observable under controlled conditions. 
We adapted this approach to provide participants with a concrete, simulated work experience through which they produced task-grounded material that could subsequently be transformed into an AI agent skill. 
This design allowed us to examine sharing decisions grounded in participants’ own work rather than in an abstract or hypothetical skill.

\subsubsection{Occupational domains selection}
Following prior experimental research on workplace AI use~\cite{noy2023experimental, reif2025evidence}, we chose four white-collar occupational domains, including \textit{marketing}, \textit{sales}, \textit{human resources}, and \textit{management}, because workers in these roles are likely to encounter AI systems that can reuse or act on professional knowledge.
~\footnote{All four roles are included in examples of AI-supported workflows, suggesting that workers in these roles are likely to use AI in their work; see \url{https://openai.com/academy/skills.}}
Moreover, these domains involve judgment-oriented tasks that could be realistically simulated within the study's time constraints.



\begin{figure}
    \centering
    \includegraphics[width=\linewidth]{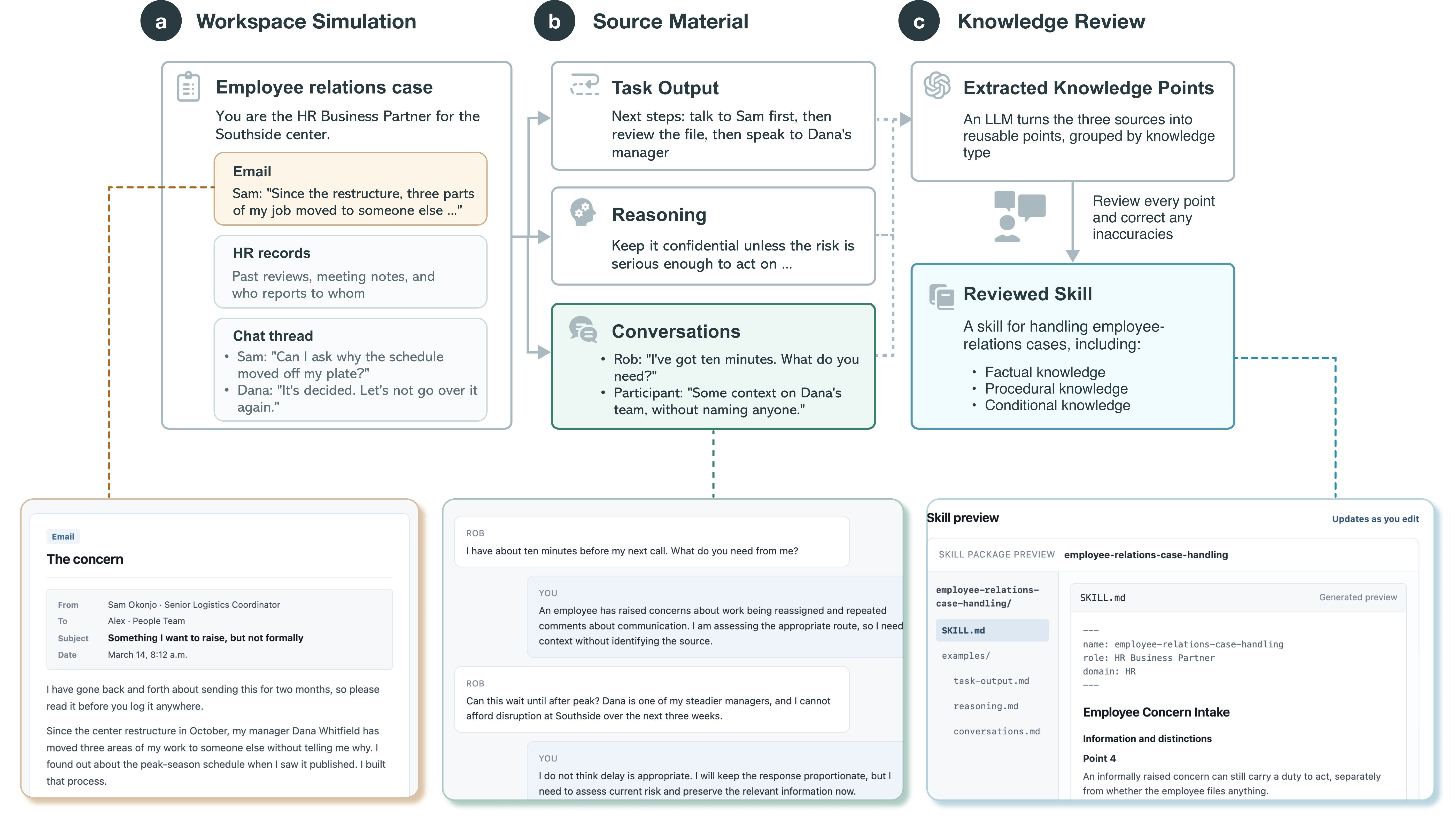}
    \caption{\textbf{Overview of the workplace simulation workflow.} The HR task is shown as an illustrative example. The upper row summarizes the three workflow stages, and the lower row shows representative interface views. (a) Participants reviewed scenario materials for an occupation-specific workplace simulation. (b) They produced three types of work artifacts: task output, reasoning, and conversations with coworkers. (c) These artifacts were synthesized into editable knowledge points and a skill preview, which participants reviewed and corrected before making sharing decisions.}
    \label{fig:screenshots}
\end{figure}

\subsubsection{Task selection and scenario development.}
To extract the skill with domain-knowledge, we designed the specific task and scenario differed by occupation. 
We selected and designed the tasks through an iterative process: 
We first collected candidate tasks and simulated workplace scenarios from prior literature and online job simulation exercises, then adapted the tasks for each occupational domain. 
The scenarios were further refined through consultations with professionals in the relevant domains. 
The task selection was guided by three criteria:
\textit{(1) occupational relevance}, grounded in task statements from an
established occupational information database\footnote{We used O*NET, a free online database of detailed occupational information: \url{https://www.onetonline.org/}.};
\textit{(2) professional expertise}, requiring occupation-specific knowledge or
judgment rather than generic administrative skills; and
\textit{(3) substantive human input}, requiring judgment beyond routine AI
assistance.\footnote{For example, we excluded tasks such as drafting standard
emails. 
For examples of AI-supported workflows across occupations, see
\url{https://openai.com/academy/skills/}.}
The four occupational simulations used in the study are summarized in Table~\ref{tab:task-design}.

\subsubsection{Simulation workflow.}
Although the simulations differed in occupational content, they shared the same elicitation structure (as shown in Figure~\ref{fig:screenshots}). 
This structure was informed by prior work on job simulations and AI-based knowledge elicitation tools~\cite{zhou2026colleague,kernan2024knowledge,kernan2023tacit}, and was designed to elicit three forms of workplace knowledge that could plausibly be extracted and reused: \textit{task product}~\cite{zhou2026colleague}, \textit{underlying reasoning}~\cite{zhou2026colleague}, and  \textit{workplace conversations}~\cite{kernan2024knowledge}. 
In each simulation, participants first reviewed a role-specific workplace scenario and described how they would handle the situation. 
They then explained the reasoning behind their decision. 
Finally, they engaged in brief conversations with two simulated coworkers, who asked what to do in the situation and why. 
These exchanges were designed to elicit additional practical knowledge and reasoning through dialogue, following prior work on conversational agents for knowledge elicitation~\cite{benderoth2025socially, kernan2023tacit}.
Task engagement statistics, such as completion time and response length, are reported in Appendix Table~\ref{tab:app-task-behavior-domain}.
Details of the design of the simulation tasks are provided in Supplementary Materials, Section 1.

\subsubsection{Validation check.}
As a validation check, the post-survey asked participants to rate the realism of the simulations and to report how often they had completed similar workplace tasks in the past year. 
Participants rated the simulations as realistic representations of their work ($M = 5.67$, $SD = 1.07$, median $= 6.00$ on a 1--7 scale). 
In addition, 77.0\% of participants with available responses reported having completed similar workplace tasks at least once in the past year. 
Full validation details are reported in the Appendix Table~\ref{tab:app-task-perception-domain}.

\begin{figure}
    \centering
    \includegraphics[width=\linewidth]{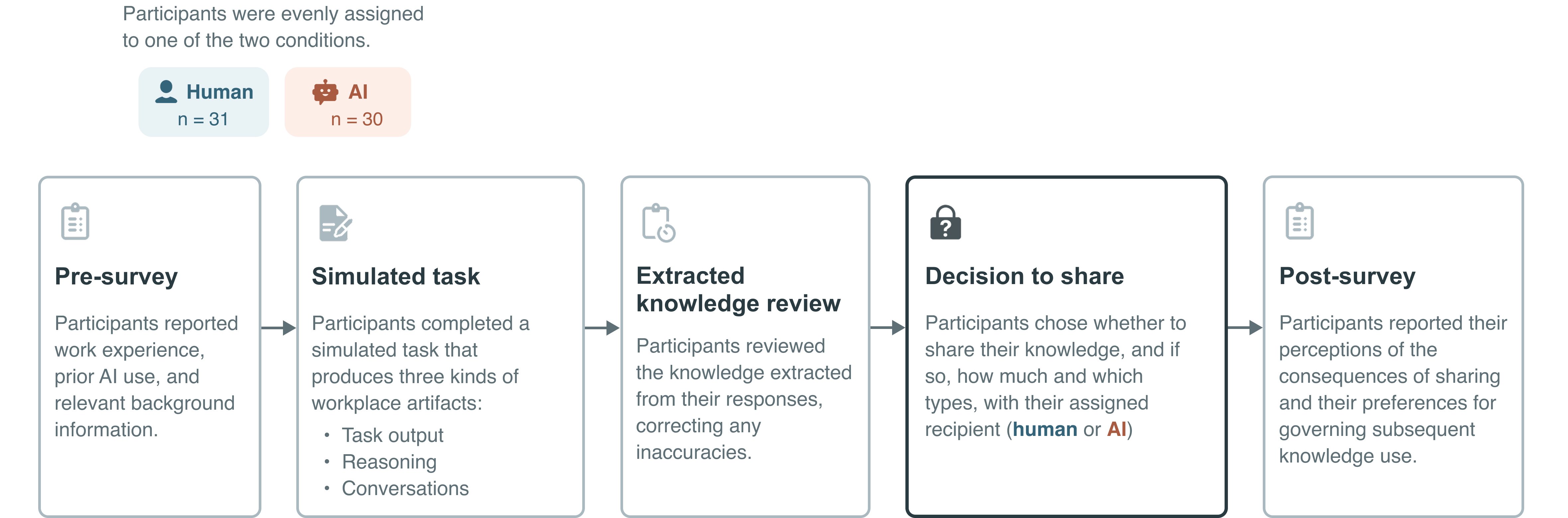}
    \caption{\textbf{Study procedure.} The study began with a pre-survey. Participants were then assigned to either the Human or AI recipient condition and completed a simulated workplace task that produced task-related artifacts. Next, they reviewed and corrected knowledge extracted from their responses. They then decided whether to share this knowledge with their assigned recipient and, if so, how much and which types to share. The study ended with a post-survey.}
    \label{fig:study-procedure}
\end{figure}

\subsection{Study Procedure}

The study procedure is illustrated in Figure \ref{fig:study-procedure}. 
Participants first completed an initial survey assessing their background, work experience, baseline attitudes toward AI, and general attitudes toward knowledge sharing.
Participants were then randomly assigned to either the Human or AI recipient condition and completed an occupational simulation. 
After completing the simulation, participants reviewed and corrected a preview of the knowledge extracted from their responses. 
Next, participants made authorization decisions about whether to share the knowledge and, if so, how much and which types to share with their assigned recipient---either \humanlabel{human professionals} or \ailabel{AI systems}.
Finally, participants completed a post-survey assessing their perceptions of the task, the extracted knowledge, and the assigned recipient, followed by a debriefing.

\subsection{System Design}
The study was administered through a custom-built web platform that participants accessed using a desktop or laptop computer. 
After completing onboarding and reading the task instructions, participants proceeded through three main stages, as illustrated in
Figure~\ref{fig:screenshots}. First, in the workspace simulation (Figure~\ref{fig:screenshots}a), participants reviewed occupation-specific source materials and completed the assigned workplace task. Second, they produced three forms of task artifacts (Figure~\ref{fig:screenshots}b): an actionable task output, an explanation of their underlying reasoning, and conversations with simulated coworkers.
Third, the platform extracted reusable knowledge points from these artifacts, which participants reviewed and corrected (Figure~\ref{fig:screenshots}c). 
After completing this workflow, participants made their knowledge-sharing decisions and received a confirmation of their selections.

\subsubsection{Frontend, Backend, and Database} The platform’s frontend was implemented using React and built with Vite. 
Participant progress was automatically synchronized with a Supabase database~\footnote{https://supabase.com/}, while local browser storage enabled recovery from accidental interruptions. 
For each session, the system recorded participants’ written outputs, conversation transcripts, extracted knowledge and subsequent revisions, sharing decisions, and relevant interaction timestamps. 
Server-side APIs handled database operations and all LLM requests.

\subsubsection{LLM Integration}
The system used \texttt{gpt-5.4-mini} via the OpenAI Responses API for two primary functions: (1) generating responses from simulated workplace coworkers (as shown in Figure~\ref{fig:screenshots}, b)  and (2) extracting and synthesizing professional knowledge from participants’ work products, reasoning, and conversations (as shown in Figure~\ref{fig:screenshots}, c). 
Prompts for the simulated coworkers specified each counterpart’s role, interaction goals, and behavioral constraints. 
Prompts for knowledge extraction instructed the model to identify discrete knowledge points and classify each as \textit{factual}, \textit{procedural}, or \textit{conditional knowledge}.

We iteratively tested and refined both prompt sets during system development.
As a check on the quality and comparability of the knowledge produced by the extraction prompts, the system presented the extracted points in an editable skill preview and allowed participants to revise or remove individual points. 
Participants generally made few revisions to the extracted knowledge points. 
Content-review behavior by condition is reported in Appendix Table~\ref{tab:app-content-review}. 
Illustrative examples of the prompts are provided in Supplementary Materials, Section 3.

\subsection{Participants}

Participants were recruited through CloudResearch~\footnote{https://www.cloudresearch.com/}. Our target population comprised college-educated, full-time professionals working in one of four target white-collar occupational groups: human resources, people management, sales, or marketing. 
The study was expected to take approximately one hour, and participants who completed the main study received a base payment of US\$10. 

When recruiting participants, we used a two-stage screening process, because CloudResearch's platform filters did not capture all of our study-specific eligibility criteria. First, we used CloudResearch's platform filters to restrict recruitment to participants who met these broad education and employment criteria. 
Second, eligible participants completed a prescreening survey assessing three study-specific criteria: (1) current employment in one of four target occupational groups; (2) prior experience performing workplace tasks similar to those used in our study; and (3) prior awareness of AI. 
Because participants were asked to evaluate an explicitly AI-enabled system, we required that they had heard of AI before the study. 
This minimal familiarity criterion follows prior HCI research~\cite{kapania2022because}.

Data collection took place from June to July 2026. Of the 113 participants who completed the prescreening survey, 65 met all eligibility criteria.
Of these, 64 entered the main study and were assigned evenly to the Human and AI recipient conditions ($n=32$ per condition). 
Three participants were subsequently excluded for failing attention checks that instructed them to select a specific response option: one from the
Human condition and two from the AI condition. 
The final analytic sample therefore comprised 61 participants (Human: $n=31$; AI: $n=30$).
Recruitment remained open throughout the planned data-collection period but was constrained by the occupation- and task-experience-specific eligibility criteria, which limited the pool of qualified participants.

The final sample comprised 34 men and 27 women, with a mean age of 34.7 years ($SD=5.9$). 
Most participants held a bachelor's degree ($n=41$) or master's degree ($n=19$), and one held a doctorate. Participants worked in management ($n=23$), sales ($n=17$), human resources ($n=13$), and marketing ($n=8$). 
Demographic and occupational characteristics were similar across the two recipient conditions. 
Full condition-specific distributions are reported in Appendix Tables~\ref{tab:app-demographics} and~\ref{tab:app-professional}.




\begin{table*}[t]
\centering
\caption{Overview of measures by research question.}
\label{tab:measure_overview}

\begingroup
\small
\setlength{\tabcolsep}{6pt}
\renewcommand{\arraystretch}{1.22}
\renewcommand{\tabularxcolumn}[1]{p{#1}}

\newcommand{\measureblock}[1]{%
    \begin{tabular}[t]{@{}
        >{\raggedright\arraybackslash}p{0.15\textwidth}
        >{\centering\arraybackslash}p{0.10\textwidth}
        >{\raggedright\arraybackslash}
        p{\dimexpr\linewidth-0.25\textwidth-4\tabcolsep\relax}
        @{}}
    #1
    \end{tabular}%
}

\begin{tabularx}{\textwidth}{
    >{\raggedright\arraybackslash}p{0.20\textwidth}
    >{\raggedright\arraybackslash}X
    @{}
}
\toprule

\multicolumn{1}{
    >{\columncolor{TableHeader}[\tabcolsep][\tabcolsep]
      \raggedright\arraybackslash}
    p{0.20\textwidth}
}{
    \textbf{RQ}
}
&
\multicolumn{1}{
    >{\columncolor{TableHeader}[\tabcolsep][0pt]
      \raggedright\arraybackslash}
    p{\dimexpr0.80\textwidth-3\tabcolsep\relax}
    @{}
}{
    \measureblock{
        \textbf{Item}
        & \textbf{Format}
        & \textbf{Description and sources}
    }
}
\\

\midrule

\textbf{RQ1:} Do knowledge workers authorize AI systems
to reuse their professional knowledge, and what do they
agree to share?
&
\measureblock{
    Overall authorization
    & Binary
    & An opt-in decision to authorize or decline knowledge
    sharing~\cite{cullen2026labor,treiman2026opting}.
    \\
    \midrule
    Authorized sharing scope
    & Checkboxes
    & Conditional on authorization, participants selected one
    or more of six categories of raw artifacts and extracted
    knowledge~\cite{ryle1945knowing}.
    \\
    \midrule
    Sharing boundaries
    & Open-ended
    & The parts of their professional knowledge that participants
    were especially willing or reluctant to include, and why.
}
\\

\midrule

\textbf{RQ2:} What benefits and risks do knowledge workers
anticipate when sharing their professional knowledge
with AI systems?
&
\measureblock{
    Anticipated consequences
    & 1--7 rating scale
    & Six items assessed (1) loss of interpretive control, (2) context-free reuse, (3) reduced expert influence, (4) uniqueness of expertise, (5) support for similar
    tasks and (6) preservation of professional knowledge~\cite{markus2001toward,kankanhalli2005contributing}.
    \\
    \midrule
    Perceived consequences
    & Open-ended
    & The anticipated benefits, risks, or other consequences
    that most influenced participants' authorization decisions,
    and why.
}
\\

\midrule

\textbf{RQ3:} What governance conditions do knowledge workers expect to increase their willingness to share professional knowledge with AI systems?
&
\measureblock{
    Governance preferences
    & 1--7 rating scale
    & Seven items assessed (1) consent, (2) control, (3) credit, (4) compensation, (5) protections concerning performance evaluation, (6) human oversight of high-stakes uses, and (7) protection against worker replacement~\cite{kyi2025governance,europeancommission2024annexiii}.
    \\
    \midrule
    Governance priorities
    & Open-ended
    & The rule or safeguard governing subsequent knowledge use that mattered most to participants' willingness to contribute, and why.
}
\\

\bottomrule
\end{tabularx}
\endgroup

\vspace{4pt}

\begin{minipage}{\textwidth}
\footnotesize
\textit{Note.} Format indicates the response format used
for each measure.
\end{minipage}

\end{table*}

\subsection{Measures}

We used a combination of closed-ended measures and open-ended questions to address the three research questions. 
Table~\ref{tab:measure_overview} summarizes the measures used to address each research question. 
Item-level results for the closed-ended RQ2 and RQ3 measures are reported in Appendix~\ref{app:measurements}, Tables~\ref{tab:app-sharing-consequences}-\ref{tab:app-contribution-conditions}.
Survey wording and response options are provided in Supplementary Materials, Section 2.

\subsubsection{Skill-Sharing Authorization (RQ1)}

\begin{figure}
    \centering
    \includegraphics[width=0.9\linewidth]{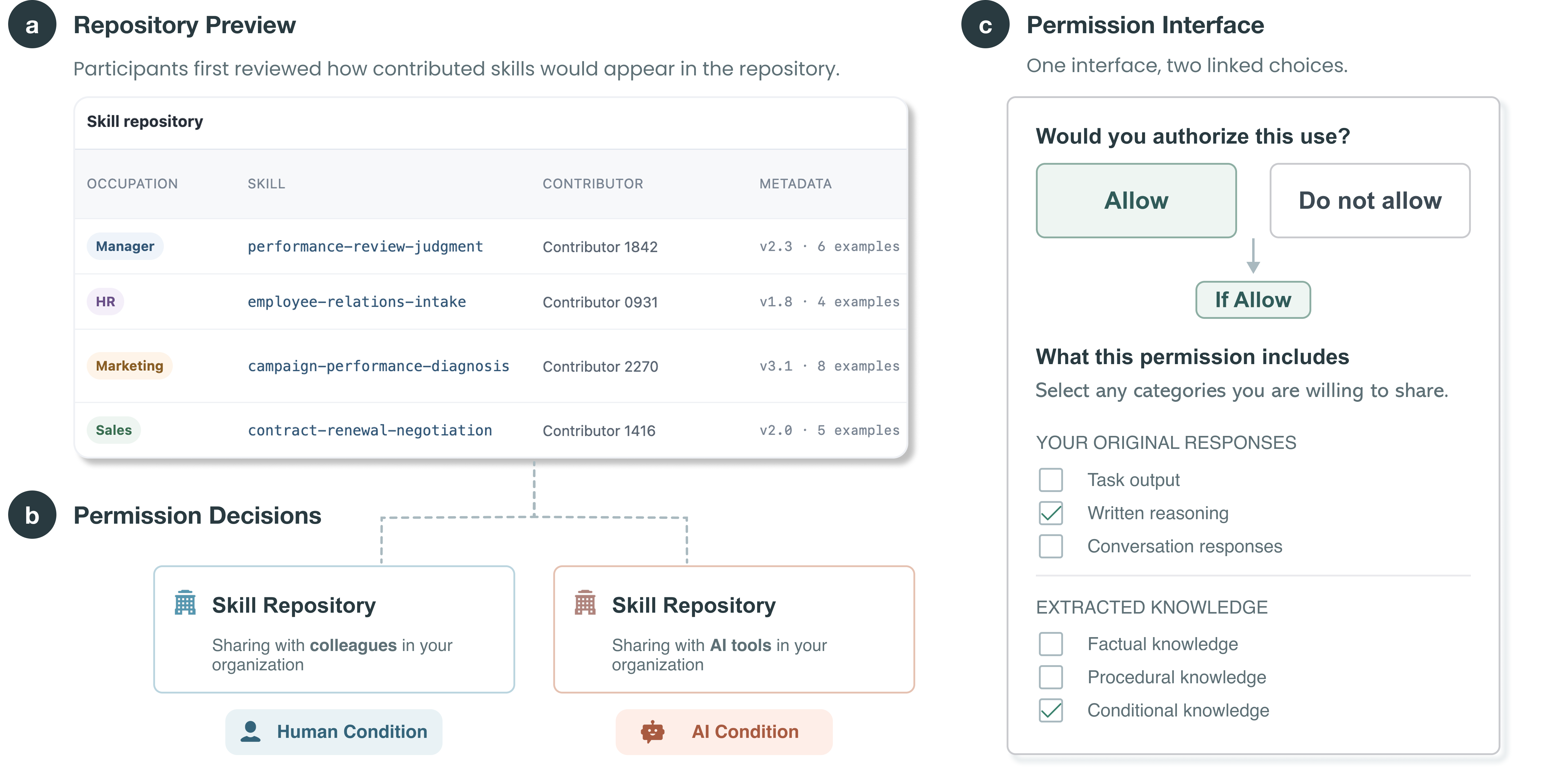}
    \caption{\textbf{Repository preview and permission decisions.} (a) Participants reviewed an illustrative skill repository to understand how material from their session would appear if shared. (b) They then made authorization decisions for an skill repository. (c) Within a single interface, participants first made an overall binary decision and, if they authorized use, selected which of six material categories to include. The checked boxes are illustrative.} 
    \label{fig:skill-repository-screenshot}
\end{figure}

We measured authorization using the repository context and interface illustrated in Figure~\ref{fig:skill-repository-screenshot}. 
Authorization was assessed at two nested levels: (1) an overall binary decision to authorize or decline sharing and, (2) conditional on authorization, checkbox selections indicating the categories of knowledge participants agreed to share. 
Following prior studies that operationalized knowledge contribution as an opt-in decision~\cite{cullen2026labor,treiman2026opting}, the overall decision was coded as 0 (\textit{did not authorize}) or 1 (\textit{authorized}). 
Participants who authorized sharing selected at least one of six categories spanning source materials produced by participants and knowledge extracted by AI.

To assess whether authorization decisions in the experimental scenario extended to a decision with real consequences~\cite{klein2019lack}, we also asked participants to decide whether to contribute their skill to an \textbf{actual repository} containing skills contributed by professionals.
Participants were informed that, if they authorized sharing, the materials they selected would be added to the repository and made accessible to [human professionals/AI systems], consistent with their assigned condition. 
They recorded this decision using the same two-level authorization interface described above.

An open-ended question complemented these measures by asking which parts of their professional knowledge they were particularly willing or reluctant to include, and why.

\subsubsection{Perceptions of Knowledge Sharing (RQ2)}

We measured participants' anticipated consequences of knowledge sharing
using six closed-ended items informed by research on knowledge reuse
\cite{markus2001toward} and social exchange
\cite{kankanhalli2005contributing}. All items followed the stem, \textit{``Sharing my knowledge with the repository would~\ldots''}
The six anticipated consequences were (1) loss of control over how one's knowledge is interpreted, (2) reuse without the necessary context, (3) reduced influence derived from one's expertise, (4) reduced uniqueness of one's expertise, (5) helping others handle similar tasks, and (6) preserving professional knowledge. 
These consequences represented three broader categories: knowledge-use risks~\cite{markus2001toward}, loss of knowledge power~\cite{kankanhalli2005contributing}, and collective benefits~\cite{kankanhalli2005contributing}.

The corresponding open-ended question asked participants which anticipated benefits, risks, or other consequences most influenced their authorization decisions and why.

\subsubsection{Governance Preferences for Knowledge Sharing (RQ3)}

We measured the conditions under which participants would be more willing to contribute their knowledge using seven items following the stem, \textit{``I would be more willing to contribute if~\ldots''} 
The items captured two complementary sets of conditions. 
The first four items were adapted from the three-C framework for governing data contributions involving intellectual property~\cite{kyi2025governance,love2024we}.  
As~\citeauthor{kyi2025governance} emphasize, however, the three Cs do not encompass all governance arrangements workers may value; consent, in particular, should be an ongoing process rather than a one-time decision.
We therefore added control as a fourth principle to capture contributors' continuing authority over their knowledge after sharing it. 
Specifically, participants rated whether they would be more willing to contribute if (1) they could give consent for specific people or groups to use their knowledge for specific purposes (consent); (2) review or correct the contributed knowledge, restrict its use, or withdraw their authorization (control); (3) receive credit when their knowledge was reused (credit); or (4) receive payment for contributing (compensation).
The remaining three items addressed workplace-specific concerns associated with AI use in employment and worker management~\cite{europeancommission2024annexiii}. 
These items assessed (5) protections against using contributed knowledge for performance evaluation, (6) requirements for qualified professional oversight of high-stakes uses, and (7) protections against using contributed knowledge as grounds for replacing workers.

The open-ended question asked participants to identify the governance rule or safeguard that mattered most to their willingness to contribute and explain why.

\subsubsection{Additional measures and data-quality checks.}
\label{sec:additional-measures}
Following previous understanding of potentail confouding variables of knowledge shairng~\cite{kankanhalli2005contributing, sulistiawan2022and, vcerne2014goes}, we collected several potential measures to check whether the participants two conditions differed on these measures that might affect the dependent variables.
These included knowledge (1) self-efficacy, (2) enjoyment in helping others, (3) generalized trust, (4) perceived organizational norms supporting knowledge sharing, and (5) organizational identification. 
All measures were collected in the pre-study survey and were used to characterize the sample and assess baseline balance.

To support data quality, we administered a comprehension check before the task. Participants selected what they would do next from four response options. To proceed, they had to correctly identify that they would complete a workplace task in a simulated setting. 
We also included attention checks that instructed participants to select a specific response option.
To assess potential demand characteristics~\cite{nichols2008good}, we administered a suspicion probe after the task, asking participants to describe what they believed the study was intended to investigate.
Participants generally described the study as examining how workplace professionals handle job-relevant tasks, suggesting that its specific focus on knowledge sharing with AI was not apparent to most participants during the simulation.
Full measures are provided in Supplementary Materials, Section 2.

\subsection{Analysis}

Our measures produced four response formats: 1--7 rating-scale responses, binary authorization decisions, checkbox selections, and open-ended responses. 
We analyzed each format using the procedures described below and integrated the quantitative and qualitative findings when interpreting each research question.
All inferential tests were two-sided, with $\alpha=.05$. 
We report effect sizes and 95\% confidence intervals alongside $p$-values where applicable.

\subsubsection{1--7 Rating-Scale Responses}

For each rating-scale outcome, we examined distributions within each recipient condition using Shapiro--Wilk tests and assessed homogeneity of
variance using median-centered Levene's tests. Because nearly all outcomes departed from normality, we compared the two independent recipient
conditions using two-sided Mann--Whitney $U$ tests. When Levene's test indicated unequal variances, we additionally conducted a Brunner--Munzel
test, which does not assume that the two distributions have the same shape.\footnote{The Brunner--Munzel and Mann--Whitney tests produced
consistent conclusions in all cases.}

\subsubsection{Binary Authorization Decisions}

Each overall authorization decision was coded as 0
(\textit{did not authorize}) or 1 (\textit{authorized}). We compared authorization rates between the two recipient conditions using two-sided Fisher's exact tests because some expected cell counts were small.

\subsubsection{Checkbox Selections}
We analyzed category selections at two levels: across the full sample and among participants who authorized organizational reuse.
First, in the full sample, each category was coded as 1 if selected and 0 otherwise, with participants who declined overall authorization coded as 0 for all categories. We compared authorization rates for each category between the AI condition and the human baseline using two-sided Fisher's exact tests. We also counted the categories released separately for written materials and extracted knowledge, yielding two counts ranging from 0 to 3 for each participant. We compared each count distribution between the AI condition and the human baseline using Fisher--Freeman--Halton exact tests.

Second, among participants who authorized organizational reuse, we examined whether release rates varied across categories within each recipient condition. We conducted separate Cochran's $Q$ tests for the three participant-authored categories and the three extracted-knowledge categories. Significant omnibus tests were followed by pairwise exact McNemar tests, with Benjamini--Hochberg correction applied across the three comparisons within each tested category group.

\subsubsection{Open-Ended Responses}

We analyzed responses to the three open-ended questions using reflexive thematic analysis. Two authors independently familiarized themselves with
the responses and generated preliminary codes. They then met regularly to compare interpretations and collaboratively develop and refine the coding
structure and candidate themes. Differences in coding were treated as opportunities for reflexive discussion rather than as errors requiring inter-rater agreement~\cite{terry2017thematic}. The authors iteratively reviewed the candidate themes against both the coded extracts and the full
dataset before defining and naming the final themes.

\section{Results}

Before conducting the primary analyses, we examined whether the two conditions differed in participant demographics or the potential confounding factors described in Section~\ref{sec:additional-measures}.
We observed no statistically significant between-condition differences. Descriptive statistics and the corresponding balance tests are reported in the Appendix~\ref{app:sample-balance}.

\begin{figure}
    \centering
    \includegraphics[width=0.7\linewidth]{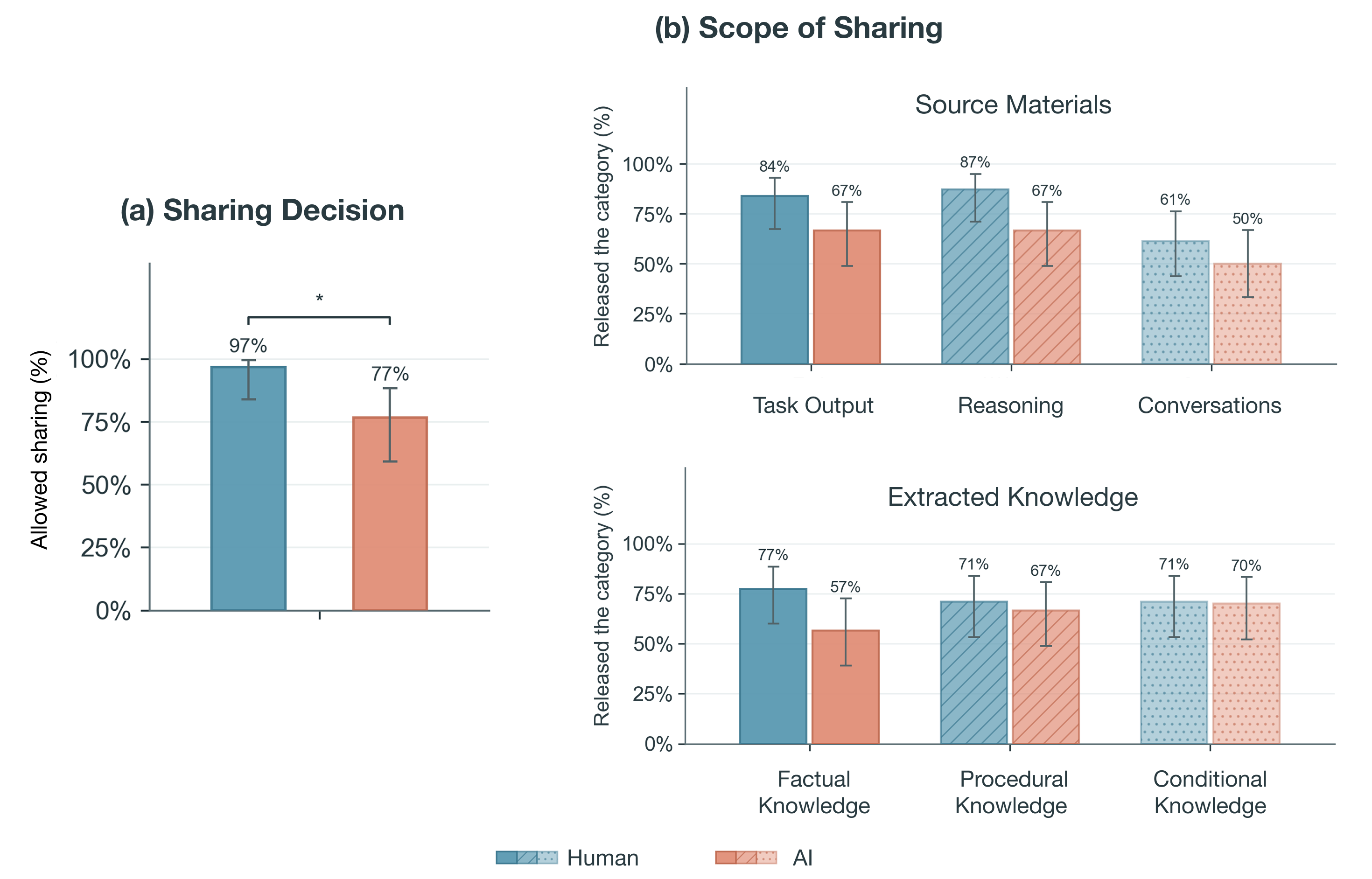}
    \caption{\textbf{Knowledge-sharing permission by assigned recipient}. (a) Percentage of participants granting permission for any reuse of their materials in the overall binary authorization decision. (b) Percentage authorizing each of six categories, grouped into written materials and extracted knowledge. Percentages use all participants in each condition; participants who declined overall authorization are counted as not releasing any category. Error bars indicate 95\% Wilson confidence intervals. The bracket marks a significant two-sided Fisher’s exact test (p < .05); no category-level comparison reached this threshold.}
    \label{fig:knowledge-sharing-permission}
\end{figure}

\subsection{Skill-Sharing Authorization (RQ1)}
To address RQ1, we examined whether participants authorized their assigned recipient to use their materials and which data and knowledge categories
they agreed to share. 
We first report the primary organizational-sharing outcomes, including the overall binary authorization decision and category-level sharing choices.
We then examine sharing patterns across material types among participants who authorized reuse, alongside the sharing boundaries they described in open-ended responses. 
Results from the corresponding public-repository decision were broadly consistent with the organizational-sharing outcomes and are reported in Appendix~\ref{app:public-repository-sharing}.



\begin{figure*}
    \centering
    \includegraphics[width=0.9\linewidth]{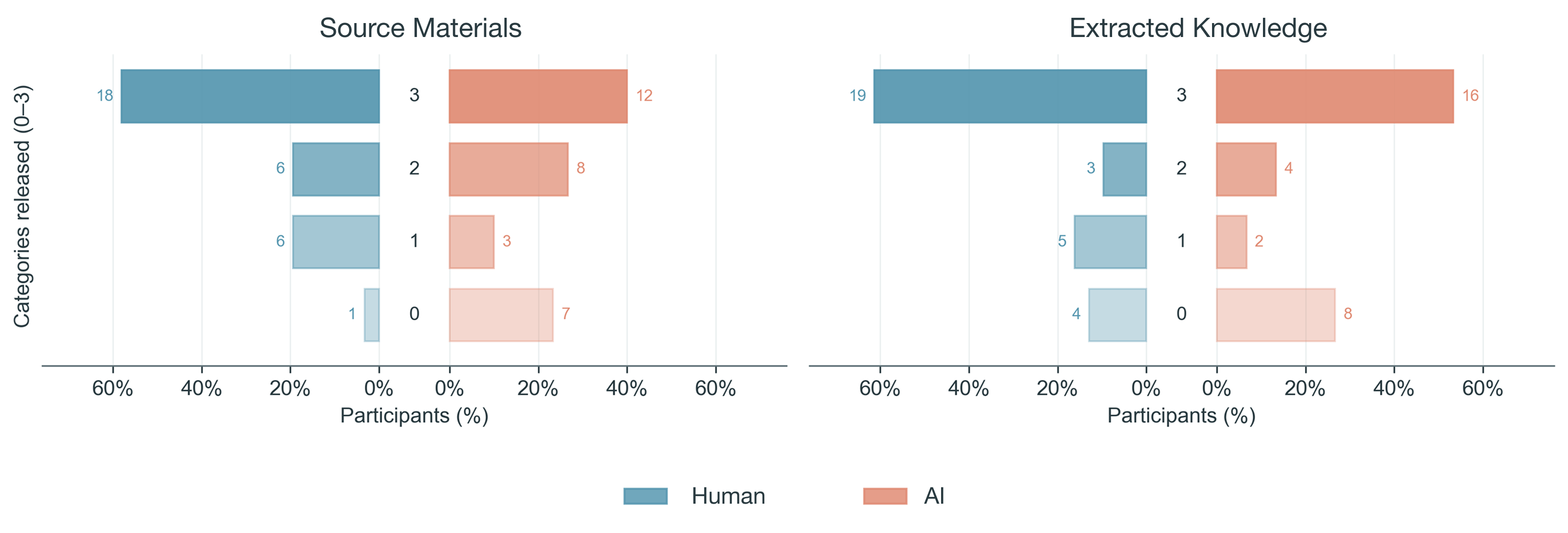}
    \caption{\textbf{Distribution of the number of categories released for organizational reuse by recipient condition}. Panels show how many of the three participant-written material categories (left) and three extracted-knowledge categories (right) participants released. Bar lengths indicate percentages of all participants in each condition, and labels report participant counts. Zero indicates that no categories in the corresponding group were released.}
    \label{fig:knowledge-sharing-counts}
\end{figure*}

\subsubsection{Overview of Skill-Sharing Decision.}
\label{sec:result-rq1-1}
As shown in Figure~\ref{fig:knowledge-sharing-permission}a, \textbf{participants were less likely to authorize AI systems to reuse any of their materials} than to authorize human professionals to do so.
Permission was granted by 23 of 30 participants in the AI condition ($76.7\%$, 95\% CI $[59\%,88\%]$), compared with 30 of 31 participants in the Human condition ($96.8\%$, 95\% CI $[84\%,99\%]$).
A two-sided Fisher's exact test indicated a significant difference, $OR=0.11$, 95\% CI $[0.02,0.96]$, $p=.026$.

Figure~\ref{fig:knowledge-sharing-permission}b presents authorization rates for each of the six data and knowledge categories. 
Participants who declined overall sharing were coded as not authorizing any individual category. 
Descriptively, \textbf{authorization rates were lower in the AI condition for all six categories}, although none of the category-level differences reached statistical significance. 
The largest difference was observed for written reasoning, which was authorized by 20 of 30 participants in the AI condition ($66.7\%$) and 27 of 31 participants in the Human condition ($87.1\%$), Fisher's exact $p=.073$. 
Thus, the category-level results were directionally consistent with the overall permission decision but did not provide conclusive evidence of differences in any specific category.

Figure~\ref{fig:knowledge-sharing-counts} shows the distribution of the number of categories participants released within each material group. 
In the AI condition, releasing no categories was descriptively more common than in the human baseline, both for participant-authored materials (23\% vs. 3\%) and extracted knowledge (27\% vs. 13\%). 
Nevertheless, releasing all three categories remained common in the AI condition (40\% and 53\%, respectively). 
The overall distributions did not differ significantly from the human baseline for either material group (Fisher–Freeman–Halton exact tests, p = .076 and .399, respectively).

\subsubsection{Types of Authorized Materials and Sharing Boundaries}
\label{sec:result-rq1-2}

\begin{figure*}
    \centering
    \includegraphics[width=\linewidth]{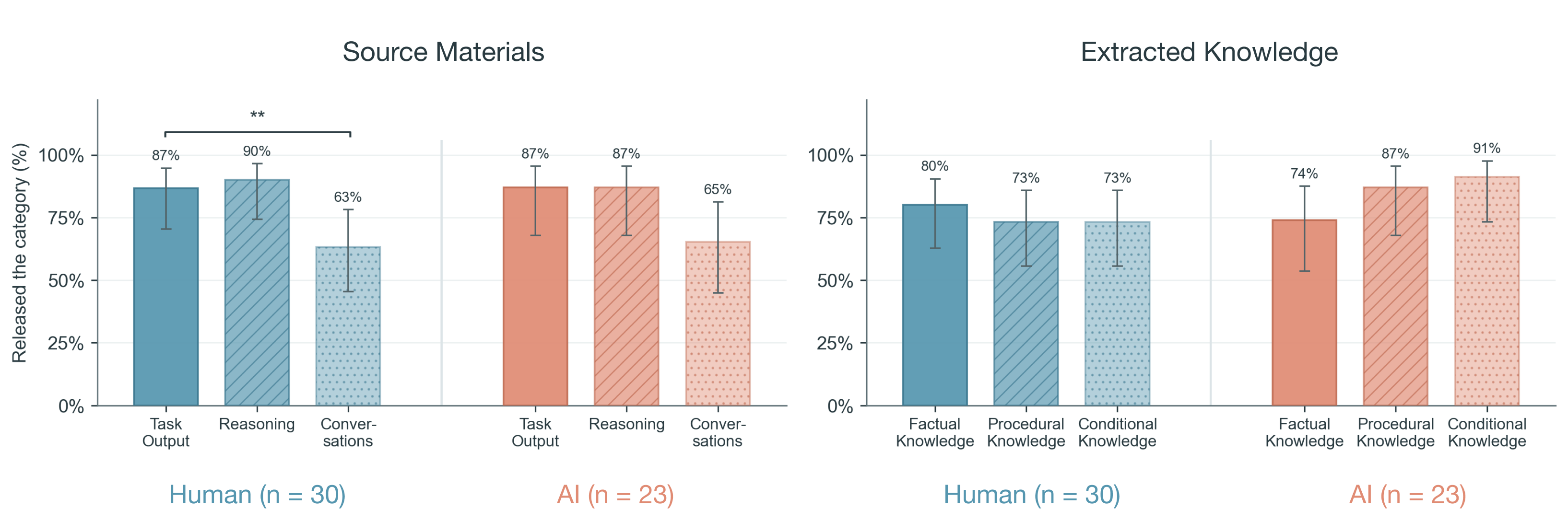}
    \caption{\textbf{Data and knowledge categories released by participants who authorized organizational reuse.} Bars show release rates by assigned recipient, with 95\% Wilson confidence intervals. The bracket indicates a significant omnibus Cochran's $Q$ test ($^{**}p<.01$); adjusted pairwise comparisons were not significant.}
    \label{fig:within-condition-category}
\end{figure*}

To examine the composition of authorized sharing, we compared category release rates within each recipient condition, restricting the analysis to participants who authorized organizational reuse (Human: $n=30$; AI: $n=23$). 
As shown in Figure~\ref{fig:within-condition-category}, among participants who authorized sharing with AI, \textbf{conversations were descriptively less frequently released} (65\%) than written reasoning or task output (87\% each), although the overall difference was not statistically significant (Cochran's $Q$, $p=.103$). 
Release rates across the three extracted-knowledge categories also did not differ significantly ($p=.115$).
The human baseline showed a similar ordering for participant-authored materials: conversations (63\%), task output (87\%), and written reasoning (90\%). Although the overall difference was significant ($p=.009$), no pairwise comparison remained significant after Benjamini--Hochberg correction. Release rates across extracted-knowledge categories did not differ significantly ($p=.607$).

Qualitative responses provided context for these descriptive patterns. When describing which materials they were unwilling to share and why, participants most often identified conversations as unsuitable for sharing (human condition: n=6, 19.4\%; AI condition: n=5, 16.7\%). 
For example, \ailabel{P28} (AI condition) expressed reservations about sharing personal feelings or private feedback about employees. Comparing conversations with more deliberate task outputs and reflections, P28 described conversations as \textit{``more unpredictable and less controlled.''} Similar concerns appeared in the Human condition: \humanlabel{P54} described conversations as subjective and liable to be taken out of context, while \humanlabel{P29} considered discussions of an employee's annual review too personal to share. These responses were consistent with the descriptive pattern of lower release rates for conversations in both conditions, suggesting that \textbf{participants' boundaries for AI reuse reflected concerns about the personal information materials revealed and the context needed to interpret them}. These concerns were also present when sharing with humans.

\subsection{Perceived Consequences of Skill-Sharing (RQ2)}
\begin{figure*}[t]
    \centering
    \includegraphics[width=\linewidth]
    {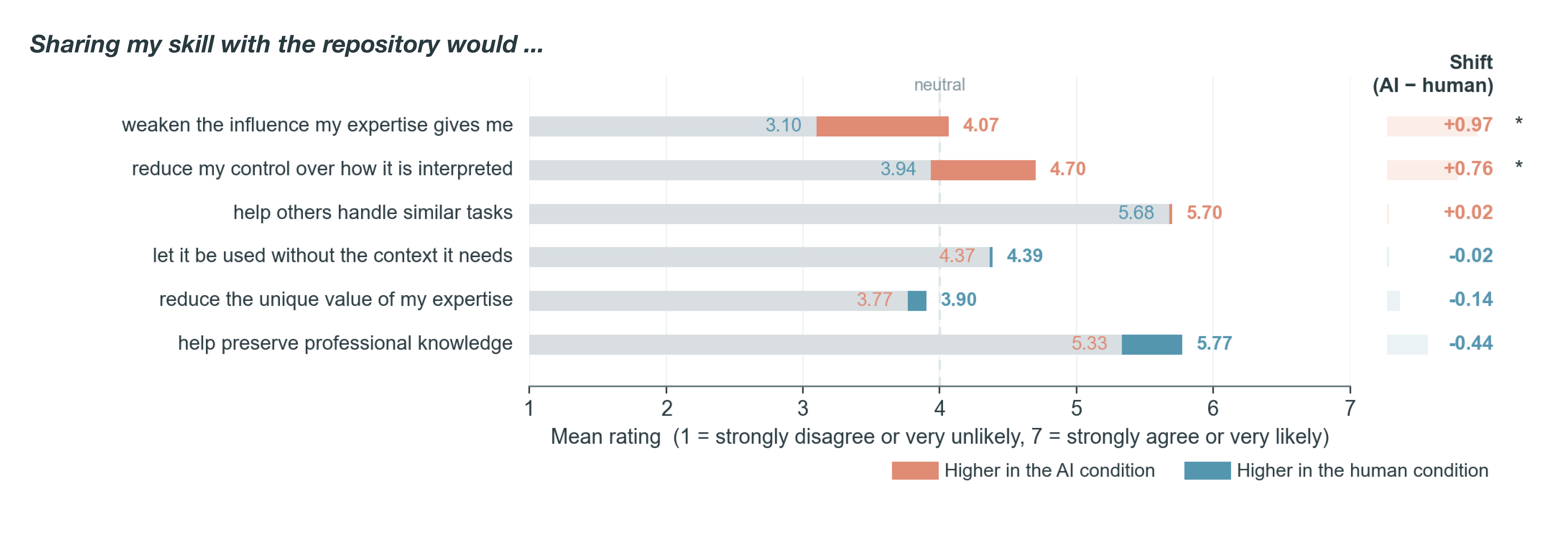}
    \caption{
    \textbf{Perceived consequences of knowledge sharing.} Participants rated six potential consequences on a 1--7 scale. Gray bars show the lower condition mean, and colored segments show the difference between conditions. Terracotta indicates higher ratings in the AI condition, whereas blue indicates higher ratings in the Human condition. The right panel reports the mean shift from Human to AI. Stars indicate two-sided Mann--Whitney comparisons ($^{*}p<.05$). }
    \label{fig:knowledge-sharing-perceptions}
\end{figure*}

To address RQ2, we examined participants' ratings of the consequences of sharing knowledge with AI systems or human professionals, summarized in Figure~\ref{fig:knowledge-sharing-perceptions}, alongside their open-ended responses about perceived benefits, risks, and concerns. 
We organize the quantitative and qualitative findings around three themes below. Full item-level statistics are reported in Appendix~\ref{app:rq2-results}.

\subsubsection{Greater Concern About Loss of Expert Influence When Sharing with AI}
\label{sec:result-rq2-1}
The largest difference in mean ratings concerned the perceived loss of expert influence when sharing with AI, while qualitative responses highlighted related concerns about professional replacement. In the survey, \textbf{participants expected sharing with AI systems to weaken the influence associated with their expertise} more than sharing with human professionals. Ratings were higher in the AI condition ($M=4.07$, $SD=1.62$) than in the Human condition ($M=3.10$, $SD=1.66$), $p=.020$, $g=0.58$, 95\% CI $[0.08,1.21]$. Ratings of reduced uniqueness of expertise did not differ significantly between conditions.

In the open-ended responses, concerns about professional replacement were concentrated in the AI condition: five participants raised the issue (16.7\%), compared with only one participant in the Human condition (3.2\%). 
These participants worried that contributing their expertise could help build a system that would subsequently replace or evaluate them. 
For example, \ailabel{P41} (AI condition) stated, \textit{``I would be more concerned about it (sharing with AI)... if it ends up replacing you because you have trained AI to replace your functions.''} 
These concerns complemented the quantitative finding about reduced expert influence by highlighting possible employment consequences of contributing expertise.

\subsubsection{Shared Decontextualization Concerns but Different Interpretive Challenges}
\label{sec:result-rq2-2}
Participants expressed concerns about how their shared knowledge would be interpreted and used outside its original context.
Quantitatively, \textbf{participants expected less control over how their knowledge would be interpreted when sharing with AI systems} ($M=4.70$, $SD=1.53$) than when sharing with human professionals ($M=3.94$, $SD=1.31$), $p=.032$, $g=0.53$, 95\% CI $[0.04,1.09]$. However, ratings of reuse without the necessary context did not differ significantly between conditions.

In the open-ended responses, decontextualization was the most common concern in both conditions, raised by 16 participants (AI: $n=7$, 23.3\%; Human: $n=9$, 29.0\%). Participants' explanations nevertheless highlighted a distinction between having sufficient contextual information and being able to interpret it appropriately. 
In the AI condition, \textbf{many participants questioned AI systems' capacity for contextual judgment itself}. For example, \ailabel{P27} described the extracted skills as suitable for \textit{``simple, straight forward conversations, but not anything that requires interpreting nuance''}. 
\ailabel{P37} emphasized that workplace situations could involve feelings and emotions, stating that \textit{``AI systems do not know empathy or compassion''} and that more complex cases should be handled by humans.
In the Human condition, the responses instead emphasized gaps in the information available to recipients and identified additional context as a way to support appropriate interpretation. 
For example, \humanlabel{P29} noted that the context needed to understand an employee rating \textit{``may not be fully expressed in the knowledge that is shared''}, while \humanlabel{P57}  requested \textit{``full guidelines and context regarding usage''}. 
These responses suggest that, although decontextualization concerned participants in both conditions, \textbf{concerns about AI reuse extended beyond the availability of context to whether the system could make appropriate judgments using it}. 
This distinction helps contextualize the higher ratings of lost interpretive control in the AI condition.

\subsubsection{Prosocial Motivations for AI Sharing Are Often Conditional}
\label{sec:result-rq2-3}

Participants in both conditions generally endorsed the potential of sharing to help others handle similar tasks and preserve professional knowledge. Survey ratings of these two benefits did not differ significantly between conditions. However, the open-ended responses revealed differences in how participants connected these benefits to their willingness to contribute.

Benefits to others were mentioned more frequently in the human condition ($n=16$, 51.6\%) than in the AI condition ($n=8$, 26.7\%). 
Human-condition participants often described helping other professionals as a direct reason to share.
\humanlabel{P10} explained, \textit{``if my method of negotiation helps someone else, I'm happy to provide pointers [...] there's no reason to gatekeep my sales tactics''}.
In the AI condition, recognizing the benefits of knowledge sharing did not necessarily translate into willingness to share through AI. 
For example, \ailabel{P27} stated, \textit{``I like being able to share my knowledge, experience, and advice with coworkers or trainees, but I'm hesitant to share with AI tools.''} 
Among the eight AI-condition participants who mentioned prosocial benefits, five explicitly endorsed the value of sharing through AI. 
Their accounts reflected \textbf{a reciprocal view in which benefiting from others' knowledge through AI motivated them to contribute their own}.
For example, \ailabel{P36} linked their contribution to the value they received from AI, explaining, \textit{``I find AI to be extremely useful when I deploy it at work and it is built on repository information gathered from various sources in the responses it provides to me.''}
They added, \textit{``I felt that everything I provided would be useful so I had no issue with setting boundaries''}.
Together, these responses suggest that \textbf{prosocial motivations extended to AI-mediated sharing, but were less frequently expressed and often accompanied by reservations or conditions}. 

\subsection{Preferred Governance of Skill Sharing (RQ3)}

\begin{figure*}[t]
    \centering
    \includegraphics[width=\linewidth]
    {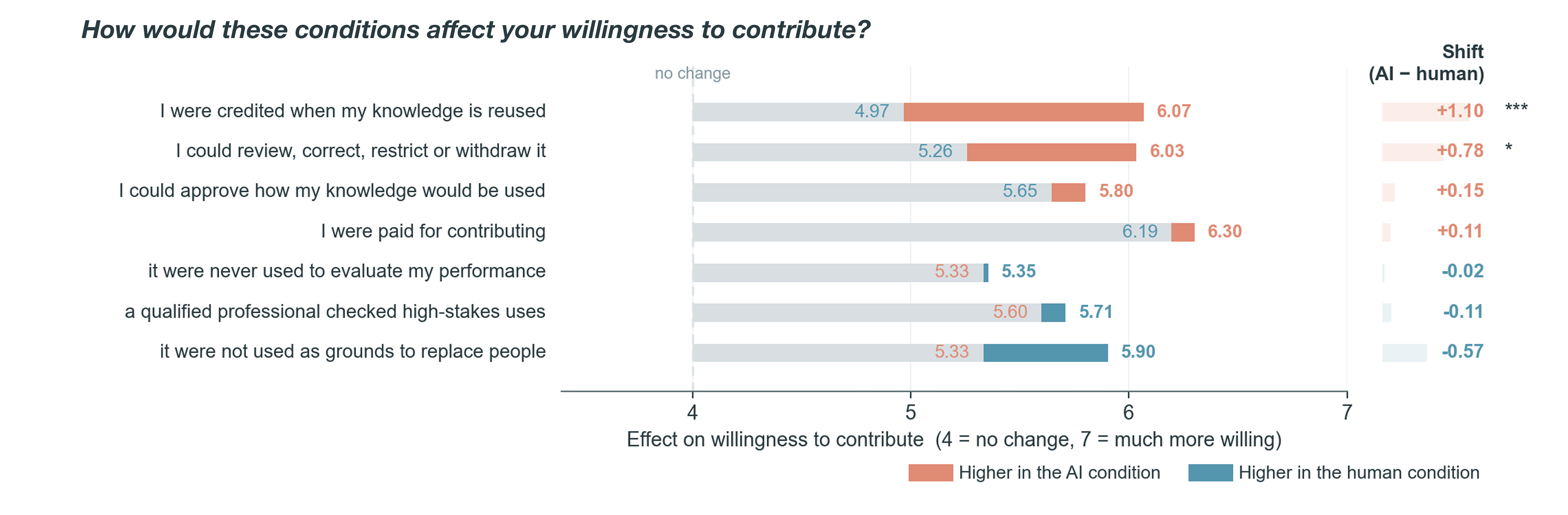}
    \caption{ 
    \textbf{Expected effects of proposed governance conditions on willingness to contribute.}
    Participants rated each condition on a 1--7 scale ($1$ = much less willing; $4$ = no change; $7$ = much more willing). Gray bars show the lower condition mean, and colored segments show the difference between conditions. Shift values are calculated as AI minus Human; terracotta indicates higher ratings in the AI condition, and blue indicates higher ratings in the Human condition. Stars indicate unadjusted two-sided Mann--Whitney comparisons ($^{*}p<.05$, $^{***}p<.001$).}
    \label{fig:perceived-encouragement}
\end{figure*}

To examine what might increase participants' willingness to share professional knowledge with AI systems or human professionals, we analyzed their ratings of seven proposed governance conditions alongside their open-ended responses about preferred rules and protections. 
As shown in Figure~\ref{fig:perceived-encouragement}, all seven conditions received mean ratings above the no-change point in both conditions, indicating that participants generally expected them to increase their willingness to contribute. 
We organize the quantitative and qualitative findings around three themes below. 
Full item-level statistics are reported in Table~\ref{tab:app-contribution-conditions} in Appendix~\ref{app:rq3-results}.

\subsubsection{Ongoing Control Over Knowledge Shared with AI}
\label{sec:result-rq3-1}
In the survey, \textbf{the ability to review, correct, restrict, or withdraw contributions was expected to increase willingness more in the AI condition} ($M=6.03$, $SD=0.89$) than in the Human condition ($M=5.26$, $SD=1.55$), $p=.022$, $g=0.60$, 95\% CI $[0.19,0.99]$. 
Consistent with this pattern, consent or contributor control was mentioned by seven participants (23.3\%) in the AI condition and only two in the Human condition (6.5\%).

Their responses further clarified what these expectations involved.
In the AI condition, participants' requests extended beyond the initial sharing decision to continued control over subsequent use. 
For example, \ailabel{P40} wanted \textit{``notifications when contributions were used, accessed, or viewed''} so that outdated information could be corrected. 
P58 similarly requested permission that \textit{``can be revoked at any time''}. 
By comparison, the two Human-condition responses focused on safeguards before sharing: 
\humanlabel{P35} wanted to \textit{``review it before submission''}, and \humanlabel{P61} requested \textit{``clear promises on data protection and explicit permission before sharing workplace information''}.
Whereas the Human-condition responses emphasized review and permission before sharing, the AI-condition responses also called for control after sharing. For these AI-condition participants, \textbf{initial consent alone was insufficient}, as they wanted to remain informed about subsequent use and retain the ability to update contributions or withdraw permission.

\subsubsection{Mixed Preferences for Credit and Anonymity When Sharing with AI}
\label{sec:result-rq3-2}
Another difference in survey ratings concerned credit.
Participants in the AI condition reported a greater expected increase in willingness when they would be credited for subsequent reuse ($M=6.07$, $SD=0.87$) than participants in the Human condition ($M=4.97$, $SD=0.80$), $p<.001$, $g=1.30$, 95\% CI $[0.72,2.18]$. 
Thus, \textbf{credit was expected to increase willingness to contribute more when knowledge was shared with AI systems} than with human professionals. 
Descriptively, however, credit received a mean rating comparable to several other governance conditions in the AI condition, whereas it was the lowest-rated condition in the Human condition.

Qualitative responses revealed differing preferences for attribution. For example, \ailabel{P40} (AI) emphasized \textit{``being credited or quoted accurately when information is reused and for what purposes''} and wanted usage notifications to enable corrections to outdated information. 
However, some AI-condition participants also emphasized anonymity; \ailabel{P36}, for example, supported reuse provided that contributions remained anonymous and sensitive information was replaced with pseudonyms. 
In the Human condition, participants rarely mentioned credit but more often expressed a preference for anonymity.
\humanlabel{P29} stated, \textit{``I was also fine sharing the things that I wrote, but I did not want my name attached,''} and P6 worried that publicizing workplace situations would invite \textit{``a lot of internal drama.''}
These responses suggest that, although credit was generally viewed favorably in the survey, its preferred form could vary across participants and should not be assumed to require publicly identifying the contributor.

\subsubsection{Protection Against Adverse Employment Consequences}
\label{sec:result-rq3-3}
Participants in both conditions generally expected protections against worker replacement and the use of contributed knowledge for performance evaluation to increase their willingness to contribute. Although neither governance condition differed significantly between the AI and Human conditions, employment-related concerns appeared in more open-ended responses in the AI condition (AI: $n=8$; Human: $n=1$). Several AI-condition participants explicitly requested guarantees that contributing knowledge would not undermine their employment. For example, \ailabel{P53} wanted assurance that the system would not \textit{``take my info \& thoughts and replace me''}. 
\ailabel{P45} similarly requested protection against \textit{``future job loss or reduction''} and the use of contributions as a basis for future evaluations. 
\ailabel{P60} called for \textit{``formal commitments that automated systems will not be used to replace human roles or evaluate performance''}. 
One similar request appeared in the Human condition, where \humanlabel{P54} wanted protection against unfair evaluation and job replacement. 
Together, these responses \textbf{framed protection from adverse employment consequences as a boundary on acceptable knowledge reuse}, with participants seeking assurances that their contributions would not be used to replace them or evaluate their performance.

\section{Discussions}


\begin{table*}[t]
\centering
\caption{Summary of findings across research questions.}
\label{tab:findings_overview}

\begingroup
\small
\setlength{\tabcolsep}{7pt}
\renewcommand{\arraystretch}{1.22}
\renewcommand{\tabularxcolumn}[1]{m{#1}}

\begin{tabularx}{\textwidth}{
    >{\raggedright\arraybackslash}m{0.28\textwidth}
    >{\raggedright\arraybackslash}X
}
\toprule

\rowcolor{TableHeader}
\textbf{Research question}
& \textbf{Findings} \\

\midrule

\textbf{RQ1:} Do knowledge workers authorize AI systems
to reuse their professional knowledge, and what do they
agree to share?
&
\begin{tabular}{@{}>{\raggedright\arraybackslash}p{\linewidth}@{}}
Participants were less likely to authorize organizational reuse
by AI systems than by human professionals (76.7\% vs.\ 96.8\%),
while differences in individual category release rates were
not statistically significant.
(Section~\ref{sec:result-rq1-1})
\\
\midrule
Among those who authorized reuse, conversations were
descriptively less often shared than task outputs or reasoning
in both conditions. Participants described concerns about
privacy and sensitive contextual information.
(Section~\ref{sec:result-rq1-2})
\end{tabular}
\\

\midrule

\textbf{RQ2:} What benefits and risks do knowledge workers
anticipate when sharing their professional knowledge
with AI systems?
&
\begin{tabular}{@{}>{\raggedright\arraybackslash}p{\linewidth}@{}}
Participants anticipated greater loss of expert influence
and interpretive control when sharing with AI.
Their accounts also raised concerns about being replaced.
(Section~\ref{sec:result-rq2-1})
\\
\midrule
Both groups worried about knowledge being taken out of context,
but AI-condition participants additionally questioned
AI's ability to interpret nuance and exercise contextual judgment.
(Section~\ref{sec:result-rq2-2})
\\
\midrule
Participants in both conditions described helping others
as a reason to share, although willingness to help did not
always translate into willingness to contribute through AI.
(Section~\ref{sec:result-rq2-3})
\end{tabular}
\\

\midrule

\textbf{RQ3:} What governance conditions do knowledge workers
expect to increase their willingness to share professional
knowledge with AI systems?
&
\begin{tabular}{@{}>{\raggedright\arraybackslash}p{\linewidth}@{}}
Credit and the ability to review, correct, restrict, or withdraw
contributions were expected to increase willingness more
in the AI condition. Participants also requested notifications
and continued control over subsequent use.
(Section~\ref{sec:result-rq3-1})
\\
\midrule
Attribution preferences varied: some participants wanted
credit, while others preferred anonymity.
(Section~\ref{sec:result-rq3-2})
\\
\midrule
Both groups valued protections against replacement and
performance evaluation. These protections appeared more often
in AI-condition accounts, although survey ratings did not
differ significantly.
(Section~\ref{sec:result-rq3-3})
\end{tabular}
\\

\bottomrule
\end{tabularx}
\endgroup

\end{table*}

Table~\ref{tab:findings_overview} summarizes our findings across the three research questions.
Together, these findings suggest that sharing expertise for AI reuse involves more than making knowledge available: workers also consider what its reuse may mean for them and what say they retain over its subsequent use.
We discuss the implications for understanding knowledge contribution and designing workplace AI systems that support workers' interests and control.

\subsection{Boundary Condition of Knowledge-Reuse Authorization for AI}

Organizations are increasingly seeking to convert workers’ situated expertise into reusable AI capabilities through agent-based knowledge-capture systems~\cite{zhou2026colleague,yuan2026ai}. These developments raise a question not fully addressed by conventional accounts of organizational knowledge sharing: does the intended recipient of workers’ knowledge matter? Our findings suggest that it does. Participants were less willing to authorize AI systems than human professionals to reuse their materials in both organizational and public settings (RQ1). However, among participants who authorized reuse, recipient type did not reliably predict which categories of material they released. 
The clearest recipient difference therefore emerged in whether reuse was permitted, rather than in fine-grained category selection after authorization.
Participants’ accounts help interpret this pattern, revealing persistent concerns about decontextualization alongside heightened concerns about the personal consequences of contributing knowledge to AI.

\subsubsection{Decontextualization as a Persistent Knowledge-Reuse Problem}
Some perceived challenges of knowledge reuse were shared across human and AI recipients, with decontextualization emerging as the most common concern in open-ended responses in both conditions (RQ2).
Participants worried that knowledge could become detached from the circumstances that made it valid and subsequently be misunderstood or misapplied. 
This concern is consistent with \textit{knowledge-reuse theory}, which emphasizes the difficulty of transferring knowledge when producers and reusers do not share the same context~\cite{markus2001toward}. 
Our findings suggest that decontextualization is not unique to AI reuse; rather, this problem remains fundamental whenever situated professional judgment is made reusable beyond its original setting.

\subsubsection{Extending Social Exchange Theory to Knowledge Contributions to AI}
Beyond the shared concern about decontextualization, participants invoked different considerations when explaining their decisions. 
Participants in the AI condition more frequently expressed professional-replacement concerns, whereas those in the Human condition more frequently invoked prosocial motivations.
AI-condition accounts also expressed prosocial motivations, often alongside reservations or a reciprocal understanding of contribution (RQ2).
This contrast extends a \textit{social-exchange perspective} on knowledge contribution to AI reuse~\cite{kankanhalli2005contributing}.
Our findings suggest that AI reuse does not displace the considerations underlying human knowledge sharing, but may change their relative importance.
Workers may recognize the collective benefits of contributing while placing greater weight on potential losses of professional value, influence, or control.
AI reuse may therefore intensify an existing tension in organizational knowledge sharing, whereby contributions that benefit others and the organization may also impose costs on the individuals who provide them.

Our findings also contribute empirical evidence to emerging accounts of workers' ``knowledge supply'' to AI, which conceptualize employee expertise as an adjustable organizational input shaped by career incentives and individual or collective ownership arrangements~\cite{cullen2026labor}. 
Whereas this work explains variation in how much workers contribute within AI knowledge-capture programs, our study compares otherwise matched human and AI recipients. 
This comparison shows that knowledge contribution cannot be understood only as variation in the quantity workers supply to AI. 
The intended recipient itself shaped authorization, while recipient type did not reliably shape category composition among those who granted permission. 
Taken together, our findings suggest that knowledge sharing with AI carries forward the contextual and social-exchange concerns of sharing among humans, while potentially intensifying tensions between the collective benefits of reuse and contributors’ own interests.

\subsection{Design Implication: Designing Provenance-Aware and Controllable Knowledge Reuse}

Based on our findings, we identify two implications for designing AI-mediated workplace knowledge reuse. 
Both concern how organizations can preserve the value of reusable knowledge without treating workers' situated expertise as an unrestricted organizational input.

\subsubsection{Preserving Context While Enabling Knowledge Reuse}
Knowledge management and CSCW research has long shown that expertise cannot simply be detached from its producer and deposited in a repository: successful reuse depends on the reuser, purpose, and context in which knowledge is applied~\cite{ackerman2013sharing, schmidt2012trouble}. 
Our findings extend this problem to AI. 
Recipient type primarily shaped whether participants authorized reuse (RQ1), but did not reliably predict which categories authorizers released. 
Nevertheless, organizational authorizers released workplace conversations less often than task outputs or written reasoning, and decontextualization was the most common unprompted concern in both conditions (RQ2, RQ3). 
AI knowledge repositories should therefore make the boundaries of reuse explicit, including the source material, whether it is retained as a raw record or transformed into an executable skill, its intended recipient and destination, and its authorized purpose.
This is especially important for agentic systems, whose value comes partly from generalizing and recombining knowledge across tasks~\cite{ling2026agent, zhou2026colleague}. 
The design goal should not be to prohibit generalization, but to distinguish adaptation within an approved domain from expansion to new purposes, populations, or consequential decisions.

\subsubsection{Making Knowledge Reuse Traceable and Contestable}
Participants expressed greater concern about losing control over how their knowledge would be interpreted or applied in the AI condition (RQ2), and their qualitative accounts called for notification, revocable permission, limits on reuse, and responsibility for downstream decisions (RQ3). 
Consistent with prior HCI work emphasizing control and ownership in AI-mediated workplace communication~\cite{kadoma2024role}, these findings suggest that transparency should support continuing influence rather than merely disclose that reuse has occurred. 
Systems should allow contributors to inspect where their knowledge is used, correct or restrict its representation, and prospectively withdraw it as deployment contexts change. Attribution can form part of this infrastructure. 
Prior HCI work has proposed richer attribution statements that distinguish contribution type, amount, initiative, and human review in human–AI co-creation~\cite{he2025contributions}. 
In our setting, visible attribution was expected to increase willingness more strongly in the AI condition (RQ3), but participants also noted that identity disclosure could create social exposure and sometimes preferred anonymity or pseudonymity. 
Systems should therefore separate internal provenance for auditing and accountability, identity disclosure to downstream users, and credit or rewards. 

\subsection{Policy Implication: Institutional Governance for AI Knowledge Management}

Beyond interface design, organizations need governance arrangements that specify who has the authority to define, modify, and enforce the conditions of knowledge reuse.
Our study allowed participants to choose whether and what to release, but workplace contributions may instead be required or automatically captured by employers, making formal consent an incomplete safeguard. 
Prior research on datafied workplaces similarly argues that individual data rights become meaningful only when supported by institutions that enable worker power and workplace democracy~\cite{dencik2024data}. 
AI knowledge reuse is therefore also a matter of organizational governance. 
Organizations should establish binding rules that limit reuse to stated purposes, involve workers or their representatives before expanding reuse to materially different purposes, assign responsibility for high-impact uses, protect contributors from undisclosed performance evaluation or adverse employment consequences, and specify how benefits generated from pooled expertise are shared. 
Participatory approaches to algorithmic management offer one model for incorporating workers’ preferences into workplace systems and decisions~\cite{lee2021participatory}. 
Such arrangements may complement contextual permissions and traceability by making reuse boundaries enforceable and giving workers a voice in defining and revising them.









\subsection{Limitation and Future Work}


First, how workers understand and approach knowledge sharing may depend on the nature of their expertise.
Although our sample covered four white-collar roles, sharing practices and concerns may differ in professions with distinct expertise requirements, regulatory constraints, or decision stakes, such as healthcare, law, and education.
Our modest sample size also limited statistical precision, particularly in the analyses restricted to participants who authorized reuse (Section~\ref{sec:result-rq1-2}).
The absence of statistically significant differences in category-level sharing should therefore not be interpreted as evidence that no differences exist.
Larger and more diverse samples are needed to estimate smaller effects and examine variation across occupations and organizational contexts.

Second, we described the AI recipient broadly to examine workers' responses to AI reuse without tying their decisions to a particular system or implementation.
This choice, however, may have bundled several properties that participants associated with AI, including scalability, autonomy, and generalization.
The study therefore cannot determine which of these properties contributed to the observed authorization gap.
It also did not distinguish among model training, retrieval-augmented generation, and agentic task execution. 
Future research should vary these technical uses and governance conditions independently to identify which characteristics most strongly shape workers’ decisions.

Lastly, although the controlled simulation enabled systematic comparisons, it could not fully reproduce the conditions of knowledge contribution in actual workplaces.
Participants may have viewed sharing as part of a paid study and therefore been more willing to contribute.
They also had explicit choices about whether and what to release, whereas organizational expectations, mandatory participation, or automatic data collection may limit such choices at work.
Contribution in practice may therefore involve greater reluctance or more complex constraints than our study captures.
Longitudinal field studies are needed to examine how organizational conditions and experiences of reuse shape workers' participation over time, alongside the quality of their contributions and the downstream uses they support.
\section{Conclusion}
This study examined how workers approach sharing their expertise for reuse by human professionals and AI systems.
In a mixed-methods study with 61 professionals, participants were less likely to authorize AI reuse than human reuse, while evidence of category-level differences was inconclusive (\textbf{RQ1}). The clearest difference therefore concerned whether workers permitted reuse at all.
Participants in both conditions raised concerns about knowledge being taken out of context, while those in the AI condition expressed greater concerns about loss of interpretive control and possible professional replacement (\textbf{RQ2}).
They also emphasized continued control over subsequent use, with individual preferences varying between receiving credit and remaining anonymous (\textbf{RQ3}).
These findings suggest that making expertise reusable through AI requires attention to the conditions under which workers are willing to contribute.
Knowledge-management systems should preserve the context and provenance of contributions, support review, correction, restrictions, and prospective withdrawal, and accommodate different attribution preferences.
Treating workers as continuing participants in decisions about how their knowledge is used can help protect their interests and support sustainable knowledge sharing with organizational AI systems.



\bibliographystyle{ACM-Reference-Format}
\bibliography{reference}

\appendix

\section{Appendix}
\subsection{Participant Characteristics and Randomization Checks}
\label{app:sample-balance}
The analytic sample contained 61 complete participants: 31 in the Human condition and 30 in the AI condition. 
Tables~\ref{tab:app-demographics}--\ref{tab:app-pre-balance} report condition-specific descriptives. 
Age was compared using a two-sided Student's independent-samples $t$ test assuming equal variances. All other numeric variables, including ordinal measures and scale composites, were compared using two-sided Mann--Whitney $U$ tests. Categorical distributions were compared using Fisher's exact tests.
Hedges' $g$ summarizes standardized mean differences for numeric variables and is signed as AI minus Human.

\subsubsection{Pre-Survey Demographic Characteristics}
Tables~\ref{tab:app-demographics} and~\ref{tab:app-professional} summarize participants' demographic and professional characteristics by condition. Across demographic characteristics, there were no statistically detectable differences between conditions (all $p \geq .401$). The same pattern held for professional background, assigned task domain, and AI experience (all $p \geq .470$). All retained participants reported full-time employment. Together, these patterns are consistent with successful allocation of the retained participants across the two recipient conditions.

\begin{center}
\small
\captionof{table}{Participant demographics by recipient condition. Continuous entries are $M$ (SD); categorical entries are $n$ (percentage among nonmissing responses).}
\label{tab:app-demographics}
\setlength{\tabcolsep}{5pt}

\begin{tabular}{p{0.38\textwidth}ccrr}
\toprule
Characteristic & Human & AI & $p$ & Effect \\
\midrule
Age, years & 35.3 (6.2) & 34.0 (5.7) & .401 & -0.21 \\
Gender &  &  & .444 & +0.11 \\
\quad Man & 19 (61.3\%) & 15 (50.0\%) &  &  \\
\quad Woman & 12 (38.7\%) & 15 (50.0\%) &  &  \\
Education &  &  & .680 & +0.14 \\
\quad Bachelor's & 22 (71.0\%) & 19 (63.3\%) &  &  \\
\quad Master's & 9 (29.0\%) & 10 (33.3\%) &  &  \\
\quad Doctorate & 0 (0.0\%) & 1 (3.3\%) &  &  \\
Race &  &  & 1.000 & +0.06 \\
\quad White & 19 (61.3\%) & 18 (60.0\%) &  &  \\
\quad Black & 6 (19.4\%) & 5 (16.7\%) &  &  \\
\quad Asian & 3 (9.7\%) & 4 (13.3\%) &  &  \\
\quad Other & 3 (9.7\%) & 3 (10.0\%) &  &  \\
Household income, USD thousands & 122.0 (60.5) & 110.2 (53.3) & .429 & -0.20 \\
\bottomrule
\end{tabular}
\end{center}

\begin{center}
\small
\captionof{table}{Professional background and assigned task domain by recipient condition. Continuous entries are $M$ (SD); categorical entries are $n$ (percentage among nonmissing responses).}
\label{tab:app-professional}
\setlength{\tabcolsep}{5pt}
\begin{tabular}{p{0.38\textwidth}ccrr}
\toprule
Characteristic & Human & AI & $p$ & Effect \\
\midrule
Eligible occupation / assigned task domain &  &  & .616 & +0.18 \\
\quad Management & 10 (32.3\%) & 13 (43.3\%) &  &  \\
\quad Sales & 11 (35.5\%) & 6 (20.0\%) &  &  \\
\quad Human resources & 6 (19.4\%) & 7 (23.3\%) &  &  \\
\quad Marketing & 4 (12.9\%) & 4 (13.3\%) &  &  \\
Industry &  &  & .470 & +0.28 \\
\quad Technology or software & 9 (29.0\%) & 5 (16.7\%) &  &  \\
\quad Financial services & 6 (19.4\%) & 4 (13.3\%) &  &  \\
\quad Healthcare & 3 (9.7\%) & 3 (10.0\%) &  &  \\
\quad Construction or real estate & 4 (12.9\%) & 2 (6.7\%) &  &  \\
\quad Education & 1 (3.2\%) & 4 (13.3\%) &  &  \\
\quad Other & 8 (25.8\%) & 12 (40.0\%) &  &  \\
Professional experience, years & 8.0 (5.4) & 7.4 (4.0) & .885 & -0.13 \\
Work AI-use frequency (1–6) & 5.39 (0.56) & 5.13 (1.01) & .509 & -0.31 \\
Agentic-AI familiarity (1–5) & 3.68 (0.94) & 3.57 (1.01) & .669 & -0.11 \\
\bottomrule
\end{tabular}
\end{center}

\subsubsection{Pre-Survey Control Variables}
No statistically significant between-condition differences were detected in the pre-task scale composites using two-sided Mann--Whitney $U$ tests. 
These variables were used to characterize the sample and assess baseline balance.

\begin{center}
\small
\captionof{table}{Pre-survey control variable scales by recipient condition. All scales range from 1 to 7; entries are $M$ (SD).}
\label{tab:app-pre-balance}
\setlength{\tabcolsep}{5pt}
\begin{tabular}{p{0.34\textwidth}ccrr}
\toprule

Characteristic & Human $M$ (SD) & AI $M$ (SD) & $p$ & $g$ \\
\midrule

\multicolumn{5}{l}{\textit{Attitudes toward AI at work}}\\
\quad Perceived humanlikeness & 2.27 (1.37) & 2.98 (1.80) & .067 & $+0.44$ \\
\quad Perceived adaptability & 5.55 (1.08) & 5.94 (0.71) & .197 & $+0.41$ \\
\quad Perceived quality of AI & 5.20 (1.25) & 5.06 (1.27) & .635 & $-0.11$ \\
\quad AI use anxiety & 2.20 (1.30) & 2.40 (1.26) & .421 & $+0.16$ \\
\quad Personal utility & 5.23 (1.27) & 5.02 (1.35) & .574 & $-0.16$ \\
\quad Job insecurity & 4.77 (1.49) & 4.88 (1.48) & .614 & $+0.07$ \\

\addlinespace

\multicolumn{5}{l}{\textit{Intrinsic motivation of knowledge sharing}}\\
\quad Knowledge-sharing enjoyment & 6.13 (0.58) & 6.13 (1.02) & .418 & $+0.01$ \\
\quad Knowledge-sharing self-efficacy & 6.28 (0.62) & 6.26 (0.66) & .973 & $-0.02$ \\

\addlinespace
\multicolumn{5}{l}{\textit{Organizational context}}\\
\quad General trust & 5.61 (1.09) & 5.59 (1.30) & .893 & $-0.02$ \\
\quad Pro-sharing norms & 6.11 (0.76) & 6.24 (0.59) & .756 & $+0.18$ \\
\quad Organizational identification & 5.42 (1.05) & 5.49 (1.26) & .606 & $+0.06$ \\

\bottomrule
\end{tabular}
\end{center}

\subsubsection{Task Exposure and Engagement}
Table~\ref{tab:app-task-behavior-domain} summarizes task exposure and engagement descriptively by occupational domain rather than by recipient condition. 
Although participants completed domain-specific tasks matched to their eligible occupations, engagement patterns were broadly similar across domains, suggesting that the task designs imposed comparable levels of workload and exposure. 
Participants spent an average of 52.32 minutes on the full study (SD = 20.58; median = 49.67) and 34.19 minutes in the task application (SD = 14.08; median = 31.60).

\begin{center}
\small
\captionof{table}{Behavioral task exposure and engagement by occupational domain. Entries are $M$ (SD).}
\label{tab:app-task-behavior-domain}
\setlength{\tabcolsep}{4pt}
\begin{tabular}{p{0.28\textwidth}cccc}
\toprule

Measure & Management & Sales & Human resources & Marketing \\
\midrule
\multicolumn{5}{l}{\textit{Time, minutes}}\\
\quad Total study completion & 51.25 (20.67) & 50.03 (15.10) & 61.90 (28.52) & 44.75 (10.89) \\
\quad Simulation task & 32.53 (13.47) & 33.29 (9.18) & 38.54 (21.17) & 33.77 (11.06) \\
\addlinespace
\multicolumn{5}{l}{\textit{Materials produced}}\\
\quad Work product (words) & 136.6 (37.6) & 140.5 (35.4) & 124.7 (15.7) & 141.0 (54.9) \\
\quad Reasoning (words) & 59.7 (21.1) & 66.4 (25.1) & 54.9 (12.4) & 55.4 (12.8) \\
\quad Conversation (words) & 152.6 (109.7) & 183.6 (87.1) & 202.6 (130.4) & 175.5 (86.1) \\
\quad Participant dialogue turns & 8.74 (0.69) & 9.18 (0.64) & 9.15 (0.90) & 9.75 (0.46) \\
\bottomrule
\end{tabular}
\end{center}

Table~\ref{tab:app-content-review} summarizes participants' editing behavior while reviewing the extracted knowledge points. 
Participants made relatively few changes overall.
The conditions were also comparable in the proportion of participants who removed or edited any point . 
Although the AI condition had a higher mean number of edited points ($p=.039$), this result was driven by only four participants and emerged among multiple uncorrected comparisons, so it should be interpreted cautiously. 
Overall, participants generally accepted the extracted knowledge without substantial revision, and content-review behavior was unlikely to explain differences in their subsequent sharing decisions.

\begin{center}
\small
\captionof{table}{Behavior at the content-review step by recipient condition. Percentages count
participants who took that action at least once; $M$ (SD) rows average over all participants,
including those who changed nothing. Fisher's exact test for percentages, Mann-Whitney for means,
uncorrected.}
\label{tab:app-content-review}
\setlength{\tabcolsep}{4pt}
\begin{tabular}{p{0.5\textwidth}ccc}
\toprule
Measure & Human & AI & $p$ \\
\midrule
Average number of knowledge points shown & 14.87 (3.18) & 14.57 (3.20)&	0.700 \\
Average number of points removed per participant & 0.55 (1.09) & 0.33 (0.71) & .681 \\
Average number of points edited per participant & 0.00 (0.00) & 0.37 (1.03) & .039 \\

\addlinespace

Percentage of participants who removed any knowledge point & 26\% (8/31) & 23\% (7/30) & 1.000 \\
Percentage of participants who edited any knowledge point & 0\% (0/31) & 13\% (4/30) & .053 \\
\bottomrule
\end{tabular}
\end{center}

\subsubsection{Post-Survey Task Evaluation}
Table~\ref{tab:app-task-perception-domain} summarizes post-task evaluations of the occupational tasks. Participants generally rated the tasks as realistic and reported moderate-to-high self-efficacy across domains. A majority of participants in each domain also reported having completed a similar task before, suggesting that the scenarios were recognizable relative to participants' professional experience.

\begin{center}
\captionof{table}{Post-task task-evaluation measures by occupational domain. Likert-scale entries are $M$ (SD); categorical entries are $n$ (percentage among available responses).}
\label{tab:app-task-perception-domain}
\centering
\small
\setlength{\tabcolsep}{4pt}
\begin{tabular}{p{0.38\textwidth}cccc}
\toprule
Measure & Management & Sales & Human resources & Marketing \\
\midrule
Perceived effort: own response (1--7) & 5.18 (0.93) & 5.14 (0.94) & 4.89 (1.76) & 4.90 (0.96) \\
Perceived effort: review extracted knowledge (1--7) & 4.93 (1.40) & 4.50 (0.76) & 4.72 (1.73) & 4.90 (1.34) \\
Post-task self-efficacy (1--7) & 5.32 (1.36) & 5.53 (0.94) & 5.31 (1.75) & 5.38 (0.92) \\
Post-task realism (1--7) & 5.50 (1.26) & 5.65 (1.00) & 5.85 (1.07) & 5.88 (0.64) \\
Similar-task frequency in the past year &  &  &  &  \\
\quad 0 & 5 (21.7\%) & 4 (23.5\%) & 3 (23.1\%) & 2 (25.0\%) \\
\quad 1--3 & 14 (60.9\%) & 4 (23.5\%) & 4 (30.8\%) & 4 (50.0\%) \\
\quad 4--10 & 2 (8.7\%) & 6 (35.3\%) & 4 (30.8\%) & 1 (12.5\%) \\
\quad More than 10 & 2 (8.7\%) & 3 (17.6\%) & 2 (15.4\%) & 1 (12.5\%) \\
\bottomrule
\end{tabular}
\end{center}

\subsection{Measurements}
\label{app:measurements}

This section reports additional results on participants’ sharing decisions for a public repository with real consequences (RQ1), along with the full item wording and item-level results for the closed-ended measures (RQ2 and RQ3).

\subsubsection{Public-Repository Sharing as an Ecological Validity Check}
\label{app:public-repository-sharing}

\begin{figure*}
    \centering
    \includegraphics[width=0.9\linewidth]{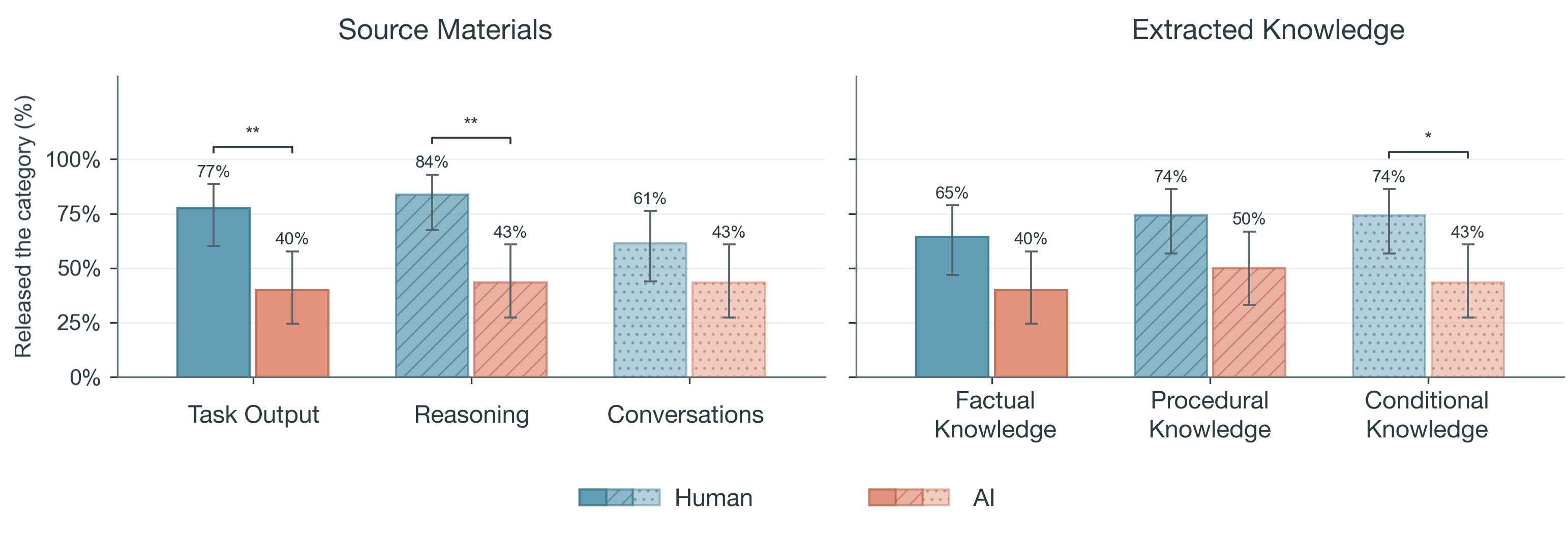}
    \caption{\textbf{Release of material categories to the public repository.} Bars show the percentage of all participants who released each category of (a) source material and (b) extracted knowledge to the public repository. Error bars show 95\% Wilson score confidence intervals. Brackets indicate significant between-condition comparisons based on two-sided Fisher's exact tests
    ($^{*}p<.05$, $^{**}p<.01$).}
    \label{fig:category-level-permission-public}
\end{figure*}

As an ecological validity check, participants also made an authorization decision for an actual public repository. 
We examined whether the effects observed for the primary organizational-sharing measure extended to this decision with real consequences.

\paragraph{Sharing permission.}
Participants assigned to the AI condition were less likely to authorize sharing through the public repository. Eighteen of 30 participants in the
AI condition granted permission ($60.0\%$, 95\% CI $[42\%,75\%]$), compared with 28 of 31 participants in the Human condition ($90.3\%$, 95\% CI $[75\%,97\%]$). A two-sided Fisher's exact test indicated a significant difference between conditions, $OR=0.16$, 95\% CI $[0.05,0.68]$, $p=.008$. This result was consistent with the primary organizational-sharing measure: in both settings, participants were less likely to authorize AI systems than human professionals to use their knowledge.

Figure~\ref{fig:category-level-permission-public} presents authorization full-sample category-level authorization rates. Participants who declined overall public sharing were coded as not releasing any of the six categories.
Authorization rates were lower in the AI condition for all six categories, consistent with the direction of the organizational-sharing results. 
For public sharing, three differences remained significant after Benjamini--Hochberg correction across the six category-level comparisons.
Participants were less likely to release written reasoning to AI systems ($13/30$, $43.3\%$) than to human professionals ($26/31$, $83.9\%$), Fisher's exact $p=.001$, $q=.008$. 
The same pattern emerged for task output (AI: $12/30$, $40.0\%$; Human: $24/31$, $77.4\%$), $p=.004$, $q=.013$, and decision rationales (AI: $13/30$, $43.3\%$; Human: $23/31$, $74.2\%$), $p=.020$, $q=.039$. 
The differences for conversations, factual knowledge, and procedural knowledge were not statistically significant after correction (all $q\geq.089$). Thus, the public-repository results followed the same overall direction as the organizational-sharing results, but the category-level differences were more pronounced.

\paragraph{Scope of sharing.}
Across the full sample, participants in the AI condition released fewer of the six knowledge categories to the public repository ($M=2.60$, $SD=2.55$, $Mdn=2$) than participants in the Human condition ($M=4.35$, $SD=2.20$, $Mdn=6$), $U=284$, $p=.006$, $g=-0.73$,
$r_{\mathrm{rb}}=-0.39$. 
The corresponding organizational-sharing comparison was in the same direction but smaller and not statistically significant. 
Thus, the public-repository result was directionally consistent with the primary outcome, although the difference in sharing scope was more pronounced when sharing had public consequences.

Among participants who authorized public sharing, no individual knowledge category differed significantly in release rate between recipient conditions (all Fisher's exact $p>.05$). 
When the recipient conditions were pooled, release rates also did not differ significantly across either the three participant-authored categories, Cochran's $Q(2)=3.70$, $p=.157$, or the three extracted-knowledge categories, $Q(2)=3.73$, $p=.155$. Because these category-level analyses were conditional on authorization, we interpret them descriptively. 
Overall, the results suggest that recipient type primarily affected whether and how much knowledge participants released, rather than systematically changing which categories they selected.

\subsubsection{Full Item-Level Results for RQ2}
\label{app:rq2-results}

Table~\ref{tab:app-sharing-consequences} reports participants' ratings of six anticipated consequences of sharing knowledge with the repository. Relative to participants in the Human condition, those in the AI condition anticipated a greater loss of influence from their expertise and less control over how their knowledge would be interpreted. We observed no statistically significant condition differences for the other four consequences.

\begin{table*}[t]
\centering
\caption{Perceived consequences of sharing knowledge with the repository.}
\label{tab:app-sharing-consequences}

\small
\setlength{\tabcolsep}{4pt}
\renewcommand{\arraystretch}{1.16}

\begin{tabularx}{\textwidth}{
    >{\raggedright\arraybackslash}X
    cccrl
}
\toprule

Item wording
& Human $M$ ($SD$)
& AI $M$ ($SD$)
& Test
& $p$
& $g$ [95\% CI] \\

\midrule

\multicolumn{6}{l}{\textit{Loss of knowledge power}}\\

\quad weaken the influence my expertise gives me
& 3.10 (1.66)
& 4.07 (1.62)
& MW
& $.020^{*}$
& $+0.58$ [$+0.08$, $+1.21$] \\

\quad reduce the unique value of my expertise
& 3.90 (1.94)
& 3.77 (2.06)
& MW
& $.707$
& $-0.07$ [$-0.58$, $+0.45$] \\

\addlinespace

\multicolumn{6}{l}{\textit{Knowledge-use risk}}\\

\quad reduce my control over how it is interpreted
& 3.94 (1.31)
& 4.70 (1.53)
& MW
& $.032^{*}$
& $+0.53$ [$+0.04$, $+1.09$] \\

\quad let it be used without the context it needs
& 4.39 (1.36)
& 4.37 (1.50)
& MW
& $.940$
& $-0.01$ [$-0.51$, $+0.49$] \\

\addlinespace

\multicolumn{6}{l}{\textit{Collective benefit}}\\

\quad help others handle similar tasks
& 5.68 (0.79)
& 5.70 (0.95)
& MW
& $.947$
& $+0.03$ [$-0.47$, $+0.53$] \\

\quad help preserve professional knowledge
& 5.77 (0.80)
& 5.33 (1.06)
& MW
& $.061$
& $-0.46$ [$-0.99$, $+0.02$] \\

\bottomrule
\end{tabularx}

\vspace{3pt}

\begin{minipage}{\textwidth}
\footnotesize
\textit{Note.} The item stem was \textit{``Sharing my knowledge with the
repository would~\ldots''}
\end{minipage}

\end{table*}

\subsubsection{Full Item-Level Results for RQ3}
\label{app:rq3-results}

Table~\ref{tab:app-contribution-conditions} reports the extent to which seven governance conditions would increase participants' willingness to contribute knowledge. Credit and control were rated as more important in the AI condition than in the Human condition. We observed no statistically significant condition differences for compensation or the three workplace-specific safeguards.

\begin{table*}[t]
\centering
\caption{Conditions increasing willingness to contribute knowledge.}
\label{tab:app-contribution-conditions}

\small
\setlength{\tabcolsep}{4pt}
\renewcommand{\arraystretch}{1.16}
\arrayrulecolor{black!55}

\begin{tabularx}{\textwidth}{
    >{\raggedright\arraybackslash}X
    cccrl
}
\toprule

Item wording
& Human $M$ ($SD$)
& AI $M$ ($SD$)
& Test
& $p$
& $g$ [95\% CI] \\

\midrule

\multicolumn{6}{l}{\textit{Consent, credit, compensation, and control}}\\

\quad I could approve how my knowledge would be used & 5.65 (0.88) & 5.80 (1.27) & MW & $.221$ & $+0.14$ [$-0.34$, $+0.72$] \\

\quad I were credited when my knowledge is reused
& 4.97 (0.80) & 6.07 (0.87) & MW & $<.001^{***}$ & $+1.30$ [$+0.72$, $+2.18$] \\

\quad I were paid for contributing
& 6.19 (0.54) & 6.30 (0.60) & MW & $.437$ & $+0.18$ [$-0.30$, $+0.72$] \\

\quad I could review, correct, restrict, or withdraw it
& 5.26 (1.55) & 6.03 (0.89) & MW & $.022^{*}$ & $+0.60$ [$+0.19$, $+0.99$] \\

\addlinespace

\multicolumn{6}{l}{\textit{Employment protection and oversight}}\\

\quad it were not used as grounds to replace people
& 5.90 (0.94) & 5.33 (1.27) & MW & $.079$ & $-0.50$ [$-0.99$, $-0.04$] \\

\quad it were never used to evaluate my performance
& 5.35 (1.33) & 5.33 (1.45) & MW & $.976$ & $-0.02$ [$-0.54$, $+0.47$] \\

\quad a qualified professional checked high-stakes uses
& 5.71 (1.07) & 5.60 (0.86) & MW & $.502$ & $-0.11$ [$-0.66$, $+0.38$] \\

\bottomrule
\end{tabularx}

\vspace{3pt}

\end{table*}

\end{document}